\documentclass[a4paper,11pt]{article}

\usepackage{jheppub}

\usepackage{amsmath,amssymb,mathtools}
\usepackage{bbold}
\usepackage{mathrsfs}
\usepackage{graphicx}
\usepackage{subfigure}
\usepackage{caption}
\usepackage{physics}
\usepackage{float}
\usepackage{array}
\usepackage{booktabs}

\newcommand{\be}{\begin{equation}}
\newcommand{\ee}{\end{equation}}
\newcommand{\bal}{\begin{align}}
\newcommand{\eal}{\end{align}}

 \arxivnumber{2504.19187}

\title{A scale invariant extension of the Georgi-Machacek model}

\author{Hamza Taibi}

\affiliation{
Department of Physics, University of M'Hamed Bougara-Boumerdes,\\
DZ-35000 Boumerdes, Algeria
}

\emailAdd{ha.taibi@univ-boumerdes.dz}

\abstract{
We present a classically scale-invariant extension of the Georgi--Machacek model, wherein
the electroweak scale is generated radiatively via the Coleman--Weinberg mechanism within
the Gildener--Weinberg framework. A real gauge-singlet scalar is introduced alongside the
usual Higgs doublet and scalar triplets, and all mass scales emerge from dimensional
transmutation. The model predicts a pseudo-Goldstone boson of scale symmetry (the scalon),
the observed 125 GeV Higgs boson, and an additional heavy neutral scalar, alongside
the charged and doubly-charged states of the original Georgi--Machacek structure. We
derive the full set of theoretical constraints from perturbative unitarity, vacuum stability,
and high-scale perturbativity, and confront the model with experimental bounds from
electroweak precision observables, LHC Higgs signal strengths, and direct LHC searches
for $H_5^{\pm\pm}$ production via vector-boson fusion with subsequent decay to
$W^\pm W^\pm$. A comprehensive numerical scan identifies viable parameter regions
consistent with all of these constraints, with every viable point developing a Landau pole
at or below $\sim 10^9$ GeV, establishing the model as a predictive effective theory valid
up to intermediate scales.
}


\begin{document}

\maketitle

\section{Introduction}

\label{intro}

The discovery of a 125 GeV Higgs boson   with properties consistent with Standard Model (SM) predictions  \cite{aad2012observation,chatrchyan2012observation} has completed the SM's particle content and underscored its extraordinary success in explaining subatomic interactions. However, the SM fails to address several key mysteries, such as the identity of dark matter, the mechanism behind neutrino masses and oscillations, and the origin of the Universe's baryon asymmetry.

A central theoretical puzzle is the hierarchy problem: the Higgs mass parameter, crucial for electroweak symmetry breaking (EWSB), is plagued by large quadratic radiative corrections from quantum loops—dominated by the top quark Yukawa coupling—that are sensitive to ultraviolet (UV) scales up to the Planck mass  \cite{susskind1979dynamics}. Absent precise cancellations, these corrections would inflate the Higgs mass far beyond the observed electroweak scale, necessitating unnatural fine-tuning.

Numerous extensions of the SM have been proposed to resolve this issue. Supersymmetry  \cite{martin199818} posits partner particles to offset quadratic divergences, but the absence of superpartners at the LHC has cast doubt on this approach. An alternative paradigm is classical scale invariance \cite{bardeen1995naturalness}, which prohibits explicit mass terms at tree level, allowing the electroweak scale to arise dynamically through dimensional transmutation driven by radiative corrections, as pioneered by Coleman and Weinberg  \cite{coleman1973radiative}. Although the minimal SM realization of this mechanism produces an unrealistically light Higgs and an unstable high-energy potential  \cite{lindner1986implications,sher1993precise}, incorporating additional scalars or gauge fields can yield stable potentials and realistic masses \cite{foot2007electroweak,espinosa2007novel,foot2008solution,iso2009classically,foot2010stable,alexander2010minimal,farzinnia2013natural,heikinheimo2014physical,karam2015dark}.

Among the well-motivated extensions that enlarge the scalar sector, the Georgi–Machacek (GM) \cite{georgi1985doubly} model stands out as a particularly attractive framework:
 it is a custodial-symmetric extension of the SM Higgs sector that supplements the complex doublet with a real triplet ($  Y=0  $) and a complex triplet ($  Y=2  $). The model's embedded custodial SU(2)$  _L \times  $ SU(2)$  _R  $ symmetry maintains the electroweak $  \rho  $ parameter at unity at tree level \cite{chanowitz1985higgs} permitting a triplet vacuum expectation value (VEV) of tens of GeV—far larger than in non-custodial triplet models. This enhanced triplet VEV enriches the phenomenology, featuring enhanced Higgs--vector boson couplings. The model further predicts a distinctive signature in the form of doubly charged Higgs decays, $H^{\pm\pm} \to W^\pm W^\pm$ \cite{falkowski2012if}, while simultaneously providing a natural framework for generating neutrino masses through the type-II seesaw mechanism \cite{cheng1980neutrino,schechter1980neutrino,mohapatra1981neutrino}.

In this work, we consider a real scalar singlet  extension of the GM model endowed with classical scale invariance. In this framework, all scalar fields acquire nonzero vacuum expectation values, which collectively trigger electroweak symmetry breaking. Owing to the intricate structure of the multi-scalar potential, we adopt the Gildener--Weinberg (GW) formalism \cite{gildener1976symmetry} to identify a flat direction along which the tree-level potential vanishes. The corresponding GW scale is then determined through renormalization-group evolution, at which one-loop quantum corrections explicitly break scale invariance, lift the vacuum degeneracy, and dynamically generate the electroweak scale.

After diagonalization of the CP-even neutral scalar mass matrix, the physical spectrum exhibits mixing among the singlet and the neutral components of the doublet and triplets. This yields a pseudo-Goldstone boson that acquires mass radiatively at one loop, alongside a  SM-like Higgs boson and an additional heavy scalar state. Importantly, the absence of trilinear scalar couplings eliminates the conventional type-II seesaw realization; instead, neutrino masses arise through a type-I seesaw mechanism involving right-handed Majorana neutrinos, whose masses are generated by the singlet VEV. This naturally points to an intermediate-scale neutrino sector with potential collider relevance.  

We establish rigorous theoretical constraints from perturbative unitarity, vacuum stability, and perturbativity at high scales through two-loop renormalization-group evolution (RGE). These are complemented by experimental limits from electroweak precision observables, LHC measurements of Higgs signal strengths, and the direct ATLAS search for $H_5^{\pm\pm}$ production via vector-boson fusion(VBF) with subsequent decay to $W^{\pm}W^{\pm}$ \cite{aad2025combination}. A thorough numerical scan identifies viable parameter regions consistent with all of these constraints, confirming the model's alignment with existing data. In addition, the renormalization-group running indicates that the model develops Landau poles at or below $\sim10^9\,\mathrm{GeV}$ throughout the viable parameter space, suggesting that it should be viewed as an effective theory valid up to intermediate scales.

The paper is structured as follows. Section 2 outlines the model construction, including the scalar potential, flat direction, mass spectrum, neutrino mass generation, and one-loop effective potential. Section 3 examines theoretical and experimental constraints. Subsequent sections detail the numerical analysis, phenomenological implications, and conclusions.

\section{The Model}
\label{sec:model}

This section presents the construction of our scale-invariant extension of the Georgi--Machacek model and derives its key properties. Employing the Coleman--Weinberg mechanism within the Gildener--Weinberg framework, we identify flat direction in the tree-level potential, compute the mass spectrum of the CP-even neutral sector, and demonstrate that a pseudo-Goldstone boson remains massless at tree level but acquires a radiatively generated mass at one loop.

\subsection{The Tree-Level Scalar Potential}
\label{subsec:potential}

The Georgi--Machacek model extends the Standard Model Higgs sector by augmenting the SM doublet $\phi$ (hypercharge $Y=1$) with a real triplet $\xi$ ($Y=0$) and a complex triplet $\chi$ ($Y=2$), arranged to preserve custodial SU(2)$_L \times$ SU(2)$_R$ symmetry. To make this symmetry explicit, we represent the fields as a bidoublet $\Phi$ and a bitriplet $\Delta$:

\begin{equation}
\Phi = \begin{pmatrix}
\phi^{0*} & \phi^+ \\
-\phi^-   & \phi^0
\end{pmatrix}, \qquad
\Delta = \begin{pmatrix}
\chi^{0*} & \xi^+  & \chi^{++} \\
\chi^-    & \xi^0  & \chi^+   \\
\chi^{--} & \xi^-  & \chi^0
\end{pmatrix}.
\label{eq:fields}
\end{equation}
The vacuum expectation values are taken to be
$\langle \Phi \rangle = \frac{v_{h}}{\sqrt{2}}\,  \mathbb{1}_{2\times2}$
and
$\langle \Delta \rangle = v_{\Delta}\, \mathbb{1}_{3\times 3}$.
These vevs are not independent, as they are constrained by the measured masses of the $W$ and $Z$ bosons, leading to
\begin{equation}
    v_{h}^2 + 8 v_{\Delta}^2 \equiv v^2 = \frac{1}{\sqrt{2} G_F} \approx (246~\mathrm{GeV})^2.
    \label{eq:vevrelation}
\end{equation}

The neutral components of the fields are parametrised as
\begin{equation}
\phi^0 = \frac{1}{\sqrt{2}} \bigl( v_h + \phi_R + i \phi_I \bigr), \qquad
\chi^0 =v_\Delta + \frac{1}{\sqrt{2}} \bigl( \chi_r + i \chi_i \bigr), \qquad
\xi^0 = \xi_r + v_\Delta.\label{fields}
\end{equation}

To impose classical scale invariance and guarantee neutrino mass generation via the type-I seesaw mechanism, we remove all dimensional parameters and introduce a real gauge-singlet scalar $  S =s+v_s$. where $v_s$ is the VEV of the singlet. For simplicity, we consider a $  \mathbb{Z}_2  $-symmetric potential where all SM fields and the singlet $  S  $ are even under the $  \mathbb{Z}_2  $ symmetry, but the bitriplet $  \Delta  $ is odd. This corresponds to the original form of the GM model as presented in the literature. The most general renormalizable, gauge- and scale-invariant scalar potential is then
\begin{align}
V_0 &= \lambda_1 \operatorname{Tr}(\Phi^\dagger \Phi)^2
    + \lambda_2 \operatorname{Tr}(\Phi^\dagger \Phi) \operatorname{Tr}(\Delta^\dagger \Delta)
    + \lambda_3 \operatorname{Tr}(\Delta^\dagger \Delta \Delta^\dagger \Delta) \notag \\
    &\quad + \lambda_4 \bigl[ \operatorname{Tr}(\Delta^\dagger \Delta) \bigr]^2
    + \lambda_5 \operatorname{Tr}\Bigl( \Phi^\dagger \tfrac{\tau^i}{2} \Phi \tfrac{\tau^j}{2} \Bigr)
               \operatorname{Tr}\Bigl( \Delta^\dagger T^i \Delta T^j \Bigr) \notag \\
    &\quad + \lambda_6 S^4 + \lambda_7 S^2 \operatorname{Tr}(\Phi^\dagger \Phi)
    + \lambda_8 S^2 \operatorname{Tr}(\Delta^\dagger \Delta),
\label{potential}
\end{align}
where $\tau^i$ are the Pauli matrices and $T^i$ are the SU(2) generators in the triplet representation:
\begin{equation}
T^1 = \tfrac{1}{\sqrt{2}} \begin{pmatrix} 0 & 1 & 0 \\ 1 & 0 & 1 \\ 0 & 1 & 0 \end{pmatrix}, \quad
T^2 = \tfrac{1}{\sqrt{2}} \begin{pmatrix} 0 & -i & 0 \\ i & 0 & -i \\ 0 & i & 0 \end{pmatrix}, \quad
T^3 = \begin{pmatrix} 1 & 0 & 0 \\ 0 & 0 & 0 \\ 0 & 0 & -1 \end{pmatrix}. 
\end{equation}

\subsection{Flat Direction}
\label{subsec:flat}

Following Gildener and Weinberg, we search for directions in field space along which the tree-level potential vanishes, consistent with the symmetry-breaking pattern SU(2)$_L \times$ U(1)$_Y \to$ U(1)$_Q$. We set all charged and CP-odd fields to zero and introduce the radial variables $h$, $ \delta$ and $ s$ as follows
\begin{equation}
\operatorname{Tr}(\Phi^\dagger \Phi) = h^2, \qquad
\operatorname{Tr}(\Delta^\dagger \Delta) = \delta^2, \qquad
S^2 = s^2.
\end{equation}
The potential then reduces to
\begin{equation}
V_0 = \lambda_1 h^4 + (\lambda_4 + \tfrac{\lambda_3}{3}) \delta^4 + \lambda_6 s^4
    + \bigl( \lambda_2 + \tfrac{\lambda_5}{2} \bigr) h^2 \delta^2
    + \lambda_7 s^2 h^2 + \lambda_8 s^2 \delta^2.
\end{equation}
We parametrize the fields along a radial direction $\varphi$ as:
\begin{equation}
h = N_h \varphi, \qquad \delta = N_\delta \varphi, \qquad s = N_s \varphi.
\label{eq:radial}
\end{equation}
where $N_i$ is a unit vector in three dimensional field space. The tree-level potential then takes the form
\begin{equation}
    V_0(\varphi, N_i) = \varphi^4\,\left(\lambda_1 N_h^4 + (\lambda_4 + \tfrac{\lambda_3}{3}) N_\delta^4 + \lambda_6 N_s^4
+ \bigl( \lambda_2 + \tfrac{\lambda_5}{2} \bigr) N_h^2 N_\delta^2
+ \lambda_7 N_s^2 N_h^2 + \lambda_8 N_s^2 N_\delta^2\right) .
\end{equation}
The flat-direction conditions require
\begin{equation}
\left. \frac{\partial V_0}{\partial N_i} \right|_{N_i = n_i} = 0 \qquad \text{and} \qquad V_0(\mathbf{n}) = 0,
\label{eq:GW_conditions}
\end{equation}
yielding the linear system
\begin{align}
2\lambda_1 n_h^2 + \bigl( \lambda_2 + \tfrac{\lambda_5}{2} \bigr) n_\delta^2 + \lambda_7 n_s^2 &= 0, \notag \\
2\lambda_6 n_s^2 + \lambda_7 n_h^2 + \lambda_8 n_\delta^2 &= 0, \notag \\
2(\lambda_4 + \tfrac{\lambda_3}{3}) n_\delta^2 + \bigl( \lambda_2 + \tfrac{\lambda_5}{2} \bigr) n_h^2 + \lambda_8 n_s^2 &= 0,
\label{mc}
\end{align}
together with the quartic constraint
\begin{equation}
\lambda_1 n_h^4 + (\lambda_4 + \tfrac{\lambda_3}{3}) n_\delta^4 + \lambda_6 n_s^4
+ \bigl( \lambda_2 + \tfrac{\lambda_5}{2} \bigr) n_h^2 n_\delta^2
+ \lambda_7 n_s^2 n_h^2 + \lambda_8 n_s^2 n_\delta^2 = 0.
\label{eq:quartic}
\end{equation}
Nontrivial solutions of the set (\ref{mc}) exist when the determinant of the coefficient matrix vanishes. Renormalisation-group evolution identifies the scale $\mu_{\rm GW}$ (the Gildener--Weinberg scale) at which this condition is satisfied.  The emergence of
this  characteristic scale in a theory that is otherwise scale-invariant is called dimensional transmutation.

The solution of the system (\ref{mc}) in terms of the quartic couplings is 
\begin{align}
n_h^2&=\dfrac{\lambda_8^2-4\lambda_6(\lambda_4 + \tfrac{\lambda_3}{3})}{\lambda_6\left(2\lambda_2+\lambda_5 - 4\left(\lambda_4 + \frac{\lambda_3}{3}\right)\right)
+ \lambda_7\left(2\left(\lambda_4 + \frac{\lambda_3}{3}\right) - \lambda_8\right)
+ \lambda_8\left(\lambda_8 - \lambda_2 - \frac{\lambda_5}{2}\right)}\notag\\
n_s^2&=\dfrac{2\lambda_7(\lambda_4 + \tfrac{\lambda_3}{3})-\lambda_8(\lambda_2+ \tfrac{\lambda_5}{2})}{\lambda_6\left(2\lambda_2+\lambda_5 - 4\left(\lambda_4 + \frac{\lambda_3}{3}\right)\right)
+ \lambda_7\left(2\left(\lambda_4 + \frac{\lambda_3}{3}\right) - \lambda_8\right)
+ \lambda_8\left(\lambda_8 - \lambda_2 - \frac{\lambda_5}{2}\right)}\notag\\
n_{\delta}^2&=\dfrac{\lambda_6 (2\lambda_2+\lambda_5)-\lambda_7 \lambda_8 }{\lambda_6\left(2\lambda_2+\lambda_5 - 4\left(\lambda_4 + \frac{\lambda_3}{3}\right)\right)
+ \lambda_7\left(2\left(\lambda_4 + \frac{\lambda_3}{3}\right) - \lambda_8\right)
+ \lambda_8\left(\lambda_8 - \lambda_2 - \frac{\lambda_5}{2}\right)}
\end{align}
we note that the $n_i$ satisy the normalization condition
\begin{equation}
    n_h^2 +  n_\delta^2 + n_s^2 = 1.
\end{equation}

\subsection{Mass Spectrum}
The CP-odd neutral and charged scalar sectors remain identical to those of the conventional Georgi–Machacek model; their mass matrices are derived in Appendix~A. Here we focus on the CP-even neutral sector.

In the custodial basis the bitriplet decomposes as \(5 \oplus 3 \oplus 1\) and the bidoublet as \(3 \oplus 1\). After absorbing the Goldstone bosons, the physical custodial states are a quintet \((H_5^{\pm\pm}, H_5^\pm, H_5^0)\), a triplet \((H_3^\pm, H_3^0)\), and two singlets \(H_1^0\) and \(\phi_R\):

\[
\begin{aligned}
H_5^{\pm\pm} &= \chi^{\pm\pm}, &
H_5^\pm &= \frac{1}{\sqrt{2}}(\chi^\pm - \xi^\pm), &
H_5^0 &= \frac{1}{\sqrt{3}}(\chi_r - \sqrt{2}\xi_r), \\
H_3^\pm &= \frac{1}{\sqrt{2}}(\chi^\pm + \xi^\pm), &
H_3^0 &= \chi_i, &
H_1^0 &= \frac{1}{\sqrt{3}}(\xi_r + \sqrt{2}\chi_r).
\end{aligned}
\]

The neutral fluctuations \(H_1^0\) and \(H_5^0\) along the custodial singlet and quintet directions satisfy $\delta^2 = (H_1^0)^2 + (H_5^0)^2$. The shifted fields after spontaneous symmetry breaking are parametrized along the flat direction as
\[
h = n_h(\varphi + v_{\varphi }),\qquad
s = n_s(\varphi  + v_{\varphi }),\qquad
\delta = n_\delta(\varphi + v_{\varphi }),
\]
where the individual VEVs are
\[
\langle h \rangle = v_h = v_{\varphi }\,n_h,\qquad
\langle s \rangle = v_s = v_{\varphi }\,n_s,\qquad
\langle \delta \rangle = v_\delta = v_{\varphi }\,n_\delta = 3v_\Delta,
\]
Here \(v_\Delta\) is the usual triplet VEV  and  $v_{\varphi }^2=v_h^2+v_s^2+v_{\delta}^2$.

Since the quintet component \(H_5^0\) carries no vacuum expectation value only the custodial singlet \(H_1^0\) mixes with the doublet neutral component \(\phi_R\) and the singlet scalar \(S\), producing the \(3\times 3\) mass-squared matrix in the basis \((\phi_R, S, H_1^0)\):
\[
\mathcal{M}_0^2 = v_\varphi^2
\begin{pmatrix}
8\lambda_1 n_h^2 & 4\lambda_7 n_s n_h & (4\lambda_2 + 2\lambda_5) n_h n_\delta \\
4\lambda_7 n_s n_h & 8\lambda_6 n_s^2 & 4\lambda_8 n_s n_\delta \\
(4\lambda_2 + 2\lambda_5) n_h n_\delta & 4\lambda_8 n_s n_\delta & 8(\lambda_4 + \frac{\lambda_3}{3}) n_\delta^2
\end{pmatrix}.
\]

Diagonalisation is performed by the rotation \(\mathcal{R}(\alpha,\beta,\gamma)\):
\[
\mathcal{R}\,\mathcal{M}_0^2\,\mathcal{R}^{-1} = \mathcal{M}_d^2,
\]
with
\[
\mathcal{R} = \begin{pmatrix}
\cos\alpha\cos\beta & -\cos\beta\cos\gamma\sin\alpha + \sin\beta\sin\gamma & -\cos\gamma\sin\beta - \cos\beta\sin\alpha\sin\gamma \\
\sin\alpha & \cos\alpha\cos\gamma & \cos\alpha\sin\gamma \\
\cos\alpha\sin\beta & -\cos\gamma\sin\alpha\sin\beta - \cos\beta\sin\gamma & \cos\beta\cos\gamma - \sin\alpha\sin\beta\sin\gamma
\end{pmatrix}.
\]
We align the flat direction with the second row of \(\mathcal{R}\) by setting
\[
\begin{aligned}
\sin\alpha &= n_h\notag\\
\cos\alpha\cos\gamma &= n_s\notag\\
\cos\alpha\sin\gamma &=n_{\delta}
\end{aligned}
\]
The matrix \(\mathcal{M}_d^2\) is then diagonal provided
\[
\tan 2\beta = \frac{(\lambda_7 - \lambda_2-\frac{\lambda_5}{2})n_s n_h n_\delta}{\lambda_1 n_h^2 - \bigl[\lambda_6 + \lambda_4 + \frac{\lambda_3}{3} -\lambda_8\bigr] n_h^2 n_\delta^2}.
\]

The resulting eigenvalues read
\[
\begin{aligned}
M_{h_1}^2 &= \frac{8\,v_\varphi^2}{1 - n_h^2}\Bigl[\cos^2\beta\,\lambda_1 n_h^2 + \sin^2\beta\,n_h^2 n_\delta^2 \bigl(\lambda_6 + \lambda_4 + \frac{\lambda_3}{3} -\lambda_8)\bigr) \\
&\qquad + \sin\beta\cos\beta\,(\lambda_7 - \lambda_2-\frac{\lambda_5}{2})n_s n_h n_\delta\Bigr], \\
M_{h_2}^2 &= 0, \\
M_{h_3}^2 &= \frac{8\,v_\varphi^2}{1 - n_h^2}\Bigl[\sin^2\beta\,\lambda_1 n_h^2 + \cos^2\beta\,n_h^2 n_\delta^2 \bigl(\lambda_6 + \lambda_4 + \frac{\lambda_3}{3} -\lambda_8)\bigr) \\
&\qquad - \sin\beta\cos\beta\,(\lambda_7 - \lambda_2-\frac{\lambda_5}{2})n_s n_h n_\delta\Bigr].
\end{aligned}
\]

The massless state \(h_2\) is the pseudo-Goldstone boson associated with classical scale symmetry (the scalon). We identify \(h_1\) with the SM-like 125\,GeV Higgs boson \(h\) and \(h_3\) with the additional heavy scalar \(H\).

\subsection{Neutrino Masses}

In the Georgi-Machacek model, small neutrino masses can naturally arise through the type-II seesaw mechanism facilitated by the complex triplet with hypercharge $  Y=2  $. However, in this scale-invariant extension, the type-II seesaw mechanism does not work because of the absence of the trilinear term in the scalar potential (e.g., a dimensionful $  \mu  $ term like $  \mu \text{Tr}(\Phi^\dagger \Delta \Phi)  $ is forbidden by classical scale invariance, preventing the induction of a sufficient triplet VEV for neutrino mass generation). To explore an alternative approach that leverages the gauge-singlet scalar $  S  $  we incorporate three right-handed Majorana neutrinos $  N_{Rj}  $ ($  j=1,2,3  $) to generate the observed neutrino masses via the type-I seesaw mechanism. This setup maintains classical scale invariance by deriving the Majorana mass scale from the singlet VEV, avoiding explicit mass terms in the Lagrangian.
We add the following Yukawa interactions to the Lagrangian:
$$\mathcal{L}_N = Y_{ij}^\nu \bar{L}_{iL} \tilde{\phi} N_{jR} + \text{h.c.} + \frac{1}{2} Y_{ij}^N S \bar{N}_{jR}^c N_{iR} + \text{h.c.},$$
where $  \tilde{\phi} = i \sigma_2 \phi^*  $, $  \phi  $ is the Higgs doublet , $  L_{iL}  $ are the left-handed lepton doublets, and $  N_{jR}  $ are the right-handed neutrino fields. For simplicity, we assume that the Majorana Yukawa matrix $  Y_{ij}^N  $ is diagonal and degenerate, $  Y^N = y_N \mathbb{I}_{3\times 3}  $.
After spontaneous symmetry breaking, the doublet acquires a VEV $  v_h  $ (where $  \langle \phi^0 \rangle = v_h / \sqrt{2}  $), and the singlet acquires $  v_s = \langle S \rangle  $. The relevant mass terms from the Lagrangian become:
$$\mathcal{L}_N^m = Y_{ij}^\nu \frac{v_h}{\sqrt{2}} \bar{\nu}_{iL} N_{jR} + \text{h.c.} + \frac{1}{2} Y_{ij}^N v_s \bar{N}_{jR}^c N_{iR} + \text{h.c.}.$$
This yields the Dirac mass matrix $  m_D = Y^\nu v_h / \sqrt{2}  $ and the Majorana mass matrix $  m_N = Y^N v_s  $. The light neutrino masses are then generated via the type-I seesaw formula:
$$m_\nu = m_D m_N^{-1} m_D^T.$$

For illustrative purposes, consider typical values where the Dirac mass is on the order of the electron mass, $  m_D \sim \mathcal{O}(0.1~\text{MeV})  $ (corresponding to small Yukawa couplings $  Y^\nu \sim 10^{-6}  $ for $  v_h \approx 246~\text{GeV}  $ in the limit of small triplet contributions). With $  y_N \sim \mathcal{O}(0.1)  $ and $  v_s \sim 1~\text{TeV}  $ consistent with viable regions identified in the parameter scans(see section 4), we obtain $  m_N \sim \mathcal{O}(100~\text{GeV})  $ and $  m_\nu \sim \mathcal{O}(0.1~\text{eV})  $, aligning with observed neutrino oscillation data.

It is important to note that the right-handed neutrino masses $  m_N  $ cannot be arbitrarily large in this scale-invariant framework. Heavy fermions contribute negatively to the one-loop effective potential in the Coleman-Weinberg mechanism, potentially destabilizing the vacuum or leading to a negative mass squared for the scalon. Thus, the type-I seesaw in this model favours intermediate-scale right-handed neutrinos, offering testable signatures at colliders like the LHC through processes such as scalon decays to neutrinos or associated production with scalars

\subsection{One-Loop Effective Potential and Scalon Mass}
\label{subsec:oneloop}

The tree-level flat direction is lifted by one-loop radiative corrections, which break scale invariance explicitly. In the $\overline{\rm MS}$ scheme the one-loop effective potential along the flat direction reads
\begin{equation}
V_1(\mathbf{n}\varphi) = A \varphi^4 + B \varphi^4 \ln \frac{\varphi^2}{\mu_{\rm GW}},
\end{equation}
with coefficients
\begin{align}
A &= \frac{1}{64\pi^2 v_\varphi^4} \Bigl\{
\operatorname{Tr} \bigl[ M_S^4 \bigl( -\tfrac{3}{2} + \log \tfrac{M_S^2}{v_\varphi^2} \bigr) \bigr]
+ 3 \operatorname{Tr} \bigl[ M_V^4 \bigl( -\tfrac{5}{6} + \log \tfrac{M_V^2}{v_\varphi^2} \bigr) \bigr] \notag \\
&\qquad - 4 \operatorname{Tr} \bigl[ M_F^4 \bigl( -1 + \log \tfrac{M_F^2}{v_\varphi^2} \bigr) \bigr] \Bigr\}, \notag \\
B &= \frac{1}{64\pi^2 v_\varphi^4} \bigl(
\operatorname{Tr} M_S^4 + 3 \operatorname{Tr} M_V^4 - 4 \operatorname{Tr} M_F^4 \bigr).
\end{align}
where $M_S$, $M_V$ and $M_F$ are the mass matrices of the scalars, vector bosons and fermions respectivly. Minimization of the full potential $V_0 + V_1$ along the flat direction gives
\begin{equation}
\ln \frac{v_\varphi}{\mu_{\rm GW}} = -\frac{1}{4} - \frac{A}{2B},
\end{equation}
so that
\begin{equation}
V_1(\mathbf{n} \varphi) = B \varphi^4 \left( \ln \frac{\varphi^2}{v_\varphi^2} - \tfrac{1}{2} \right).
\end{equation}
The scalon mass is then obtained from the second derivative:
\begin{equation}
m_s^2 = \left. \frac{\partial^2 V_1}{\partial \varphi^2} \right|_{\varphi = v_\varphi} = \frac{1}{8\pi v_\varphi^2} \bigl(
5 m_{H_5}^4 + 3 m_{H_3}^4 + m_h^4 + m_H^4 + 6 m_W^4 + 3 m_Z^4 - 12 m_t^4 -6 m_N^4 \bigr).
\end{equation}
Positivity of $m_s^2$ requires the scalar and gauge boson contributions to overcome the large negative top-quark and right handed neutrino terms:
\begin{equation}
5 m_{H_5}^4 + 3 m_{H_3}^4 + m_h^4 + m_H^4 + 6 m_W^4 + 3 m_Z^4 > 12 m_t^4+6 m_N^4.
\end{equation}

Thus far, we have presented the theoretical foundation of the scale-invariant GM model (SIGM), including its scalar potential, symmetry-breaking pattern, and one-loop mass generation. To ensure the model’s viability, we must now subject it to a suite of theoretical and experimental constraints. In the next section, we examine bounds from tree-level unitarity, vacuum stability, renormalization group evolution, electroweak precision observables, and Higgs coupling measurements.

\section{Theoretical and Experimental Constraints}

\subsection{Tree-Level Unitarity}

Perturbative unitarity of scalar field scattering amplitudes requires that the zeroth partial wave amplitudes, \(a_0\), satisfy \(|a_0| \leq 1\) (or equivalently, \(\Re(a_0) \leq \frac{1}{2}\)). Unitarity conditions for the scalar singlet extension of the GM model have been thoroughly investigated in \cite{campbell2017real}. Here, we recast those results using our notation:
\begin{align}
8\pi >\; &\left| 12\lambda_1 + 14\lambda_3 + 22\lambda_4 \pm \sqrt{(12\lambda_1 - 14\lambda_3 - 22\lambda_4)^2 + 144\lambda_2^2} \right|, \nonumber \\
8\pi >\; &\left| 4\lambda_1 - 2\lambda_3 + 4\lambda_4 \pm \sqrt{(4\lambda_1 + 2\lambda_3 - 4\lambda_4)^2 + 4\lambda_5^2} \right|, \nonumber \\
8\pi >\; &|16\lambda_3 + 8\lambda_4|, \nonumber \\
8\pi >\; &|4\lambda_3 + 8\lambda_4|, \nonumber \\
8\pi >\; &|4\lambda_2 - \lambda_5|, \nonumber \\
8\pi >\; &|4\lambda_2 + 2\lambda_5|, \nonumber \\
8\pi >\; &|4\lambda_2 + 4\lambda_5|, \nonumber \\
8\pi >\; &|4\lambda_7|, \nonumber \\
8\pi >\; &|4\lambda_8|, \nonumber \\
\lambda_6 <\; &\frac{1}{6}\left(4\pi + \frac{2\lambda_7^2(7\lambda_3 + 11\lambda_4 - \pi) + 9\lambda_8^2(3\lambda_1 - \pi) - 18\lambda_2\lambda_7\lambda_8}{2(7\lambda_3 + 11\lambda_4 - \pi)(3\lambda_1 - \pi) - 9\lambda_2^2}\right), \nonumber \\
\lambda_6 >\; &\frac{1}{6}\left(-4\pi + \frac{2\lambda_7^2(7\lambda_3 + 11\lambda_4 + \pi) + 9\lambda_8^2(3\lambda_1 + \pi) - 18\lambda_2\lambda_7\lambda_8}{2(7\lambda_3 + 11\lambda_4 + \pi)(3\lambda_1 + \pi) - 9\lambda_2^2}\right).
\label{eq:uniconstr}
\end{align}
\subsection{Vacuum Stability}
For the vacuum to be stable, the scalar potential must be bounded from below (BFB) in every direction of field space. To derive the necessary conditions for stability, we adopt the parametrization introduced in Ref \cite{arhrib2011higgs}. In this approach, the scalar fields are expressed in terms of a radial variable and two angular coordinates. Specifically, we define
\begin{equation}
r = \sqrt{\mathrm{Tr}(\Phi^\dagger\Phi) + \mathrm{Tr}(\Delta^\dagger\Delta) + S^2},
\end{equation}
\begin{equation}
r^2 \cos^2 \alpha_2\, \sin^2 \alpha_1 = \mathrm{Tr}(\Delta^\dagger\Delta),\quad
r^2 \sin^2 \alpha_2\, \sin^2 \alpha_1 = \mathrm{Tr}(\Phi^\dagger\Phi),\quad
r^2 \cos^2\alpha_1 = S^2,
\end{equation}
with \(r \in [0,\infty]\), \(\alpha_1 \in [0,\pi/2]\), and \(\alpha_2 \in [0,\pi/2]\).
We further introduce the orbit parameters $  \zeta  $ and $  \omega  $, characteristic of the Georgi--Machacek model, as defined in Ref.~\cite{hartling2014decoupling}:
\be
\zeta = \frac{\mathrm{Tr}(\Delta^\dagger\Delta\,\Delta^\dagger\Delta)}{\mathrm{Tr}(\Delta^\dagger\Delta)^2},\quad
\omega = \frac{\mathrm{Tr}\Bigl(\Phi^\dagger \frac{\tau_i}{2}\Phi\,\frac{\tau_j}{2}\Bigr)\,\mathrm{Tr}(\Delta^\dagger T_i\Delta T_j)}{\mathrm{Tr}(\Phi^\dagger\Phi)\,\mathrm{Tr}(\Delta^\dagger\Delta)}.
\ee
It was shown in ~\cite{hartling2014decoupling} that $  \zeta \in [1/3,1]  $ and $  \omega \in [-1/4,1/2]  $.
By substituting $  y = \sin^2\alpha_1  $ and $  x = \cos^2\alpha_2  $, the potential $  V_0(r,\alpha_1,\alpha_2)  $ can be recast as
\be
V_0 = y^2 \Bigl[\lambda_1 (1-x)^2 + (\lambda_4 + \lambda_3 \zeta)x^2 + (\lambda_2 + \lambda_5 \omega)x(1-x)\Bigr] + \lambda_6 (1-y)^2 + y(1-y)\Bigl[\lambda_7 (1-x) + \lambda_8 x\Bigr].
\ee
To guarantee that the potential remains bounded from below, we examine a generic biquadratic function
$$f(z) = A z^2 + B (1-z)^2 + C z (1-z),$$
which is positive for all $  z \in [0,1]  $ provided that
\be
A > 0,\quad B > 0,\quad C + 2\sqrt{AB} > 0.
\ee
In our case, we identify
\be
A = \lambda_1 (1-x)^2 + (\lambda_4 + \lambda_3 \zeta)x^2 + (\lambda_2 + \lambda_5 \omega)x(1-x),\quad B = \lambda_6,
\ee
\be
C = \lambda_7 (1-x) + \lambda_8 x,
\ee
and require
\be
\lambda_7 (1-x) + \lambda_8 x > -2\sqrt{AB}.\label{TC}
\ee
This identification leads to a set of conditions on the quartic couplings. The analysis of the first condition (i.e., $  A > 0  $) has been carried out in Ref.~\cite{hartling2014decoupling} and yields the following BFB constraints:
\be
\lambda_1 > 0,\quad \lambda_4 >
\begin{cases}
-\frac{1}{3}\lambda_3 & \text{if } \lambda_3 \geq 0,\\
-\lambda_3 & \text{if } \lambda_3 < 0,
\end{cases}\label{l1c}
\ee
\be
\lambda_2 >
\begin{cases}
-\frac{1}{2}\lambda_5 - 2\sqrt{\lambda_1 \Bigl(\frac{1}{3}\lambda_3 + \lambda_4\Bigr)} & \text{if } \lambda_5 \leq 0 \text{ and } \lambda_3 \geq 0,\\
-\omega_+(\zeta)\lambda_5 - 2\sqrt{\lambda_1 \Bigl(\zeta\lambda_3 + \lambda_4\Bigr)} & \text{if } \lambda_5 \leq 0 \text{ and } \lambda_3 < 0,\\
-\omega_-(\zeta)\lambda_5 - 2\sqrt{\lambda_1 \Bigl(\zeta\lambda_3 + \lambda_4\Bigr)} & \text{if } \lambda_5 > 0,
\end{cases}\label{sc}
\ee
with
$$\omega_\pm(\zeta) = \frac{1}{6}(1-B) \pm \frac{\sqrt{2}}{3}\sqrt{(1-B)\Bigl(\frac{1}{2}+B\Bigr)},\quad B = \sqrt{\frac{3}{2}\Bigl(\zeta - \frac{1}{3}\Bigr)} \in [0,1].$$ The second condition $B>0$ leads to 
\be
\lambda_6>0\label{6c}
\ee
The third condition (\ref{TC}), involving the mixed term, requires a more detailed analysis and naturally divides into two cases:
\begin{enumerate}
\item \textbf{Case 1:} If $  \lambda_7 > 0  $ and $  \lambda_8 > 0  $, the inequality is automatically satisfied provided that the previous constraints ensure $  AB > 0  $.
\item \textbf{Case 2:} When $  \lambda_7 < 0  $ and/or $  \lambda_8 < 0  $, further distinctions arise:
\begin{enumerate}
\item \textbf{Scenario 1:} The requirement $  C > 2\sqrt{AB}  $ leads to a contradiction because it would imply both $  \lambda_7  $ and $  \lambda_8  $ are positive, a situation already covered by Case 1.
\item \textbf{Scenario 2:} When $  -2\sqrt{AB} < C < 2\sqrt{AB}  $, one obtains the inequality
\be
\begin{aligned}
&\Bigl(4\lambda_1\lambda_6 - \lambda_7^2\Bigr)(1-x)^2 + \Bigl(4\lambda_6\bigl(\lambda_4+\lambda_3\,\zeta\bigr)-\lambda_8^2\Bigr)x^2\\[1mm]
&\quad + \Bigl(4\lambda_6\Bigl(\lambda_2+\omega\lambda_5\Bigr) - 2\lambda_7\lambda_8\Bigr)x(1-x) > 0.
\end{aligned}
\ee
Imposing the copositivity condition on this expression yields the following relations:
\be
4\lambda_1\lambda_6 - \lambda_7^2 > 0,\quad 4\lambda_6\bigl(\lambda_4+\lambda_3\,\zeta\bigr)-\lambda_8^2 > 0,
\ee
and
\be
4\lambda_6\Bigl(\lambda_2+\omega\lambda_5\Bigr) - 2\lambda_7\lambda_8 + 2\sqrt{\Bigl(4\lambda_1\lambda_6 - \lambda_7^2\Bigr)\Bigl(4\lambda_6\bigl(\lambda_4+\lambda_3\,\zeta\bigr)-\lambda_8^2\Bigr)} > 0.\label{TC2}
\ee
These requirements further constrain the parameters:
\be
-2\sqrt{\lambda_1 \lambda_6} < \lambda_7 < 2\sqrt{\lambda_1 \lambda_6},\label{inq3}
\ee
\be
\begin{cases}
-2\sqrt{\lambda_6 \Bigl(\lambda_4 + \frac{\lambda_3}{3}\Bigr)} < \lambda_8 < 2\sqrt{\lambda_6 \Bigl(\lambda_4 + \frac{\lambda_3}{3}\Bigr)} & \text{if } \lambda_3 \geq 0,\\
-2\sqrt{\lambda_6 (\lambda_4 + \lambda_3)} < \lambda_8 < 2\sqrt{\lambda_6 (\lambda_4 + \lambda_3)} & \text{if } \lambda_3 < 0.
\end{cases}\label{inq4}
\ee
Additionally, the relation (\ref{TC2}) imposes
\be
\lambda_2 >
\begin{cases}
-\frac{1}{2}\lambda_5 + \frac{\lambda_7 \lambda_8}{2\lambda_6} - \sqrt{\Bigl(\lambda_1 - \frac{\lambda_7^2}{4\lambda_6}\Bigr)\Bigl(\frac{1}{3}\lambda_3+\lambda_4-\frac{\lambda_8^2}{4\lambda_6}\Bigr)} & \text{if } \lambda_5 \leq 0,\;\lambda_3 \geq 0,\\
-\omega_+(\zeta)\lambda_5 + \frac{\lambda_7 \lambda_8}{2\lambda_6} - \sqrt{\Bigl(\lambda_1 - \frac{\lambda_7^2}{4\lambda_6}\Bigr)\Bigl(\zeta\lambda_3+\lambda_4-\frac{\lambda_8^2}{4\lambda_6}\Bigr)} & \text{if } \lambda_5 \leq 0,\;\lambda_3 < 0,\\
-\omega_-(\zeta)\lambda_5 + \frac{\lambda_7 \lambda_8}{2\lambda_6} - \sqrt{\Bigl(\lambda_1 - \frac{\lambda_7^2}{4\lambda_6}\Bigr)\Bigl(\zeta\lambda_3+\lambda_4-\frac{\lambda_8^2}{4\lambda_6}\Bigr)} & \text{if } \lambda_5 > 0.
\end{cases}\label{inq5}
\ee
\end{enumerate}
\end{enumerate}
Collectively, the inequalities (\ref{l1c}), (\ref{sc}),(\ref{6c}), (\ref{inq3}), (\ref{inq4}), and (\ref{inq5}) form the complete set of conditions required to ensure that the scalar potential remains bounded from below, thereby guaranteeing vacuum stability.

\subsection{Renormalization Group Evolution}

To test the validity of our model at high energy scales, we calculate the
renormalization group evolution of the quartic couplings and impose the
perturbativity condition
\begin{equation}
|\lambda_i| <4 \pi
\end{equation}
up to the Planck scale ($M_P$).

The RGE of the quartic couplings in the Georgi--Machacek model is complicated
by the fact that custodial symmetry is broken at the loop level by the
hypercharge interaction, as first noted in Ref  \cite{gunion1991naturalness}.
This custodial-symmetry-breaking effect can grow large at higher energy scales
and therefore cannot be neglected. Consequently, we follow the approach of
\cite{blasi2017effects}, where the most general custodial-symmetry-breaking
potential is used to compute the RGE.

For the scale-invariant GM model, the most general gauge and scale-invariant
scalar potential with custodial symmetry breaking is given by
\begin{align}
V(\phi,\chi,\xi,S) &= 
\lambda (\phi^{\dagger} \phi)^2
+ \rho_1 \bigl[\Tr(\chi^{\dagger} \chi)\bigr]^2
+ \rho_2 \Tr(\chi^{\dagger} \chi \chi^{\dagger} \chi)
+ \rho_3 \Tr(\xi^4)
\nonumber \\
&\quad
+ \rho_4 \Tr(\chi^{\dagger} \chi)\Tr(\xi^2)
+ \rho_5 \Tr(\chi^{\dagger} \xi)\Tr(\xi \chi)
+ \sigma_1 \Tr(\chi^{\dagger} \chi)\,\phi^{\dagger} \phi
\nonumber \\
&\quad
+ \sigma_2 \phi^{\dagger} \chi \chi^{\dagger} \phi
+ \sigma_3 \Tr(\xi^2)\,\phi^{\dagger} \phi
+ \sigma_4 \left( \phi^{\dagger} \chi \xi \phi^c + \text{h.c.} \right)
\nonumber \\
&\quad
+ \lambda_6 S^4
+ \lambda_7 S^2 \phi^{\dagger} \phi
+ \lambda_{8a} S^2 \Tr(\chi^{\dagger} \chi)
+ \lambda_{8b} S^2 \Tr(\xi^2),
\label{pot_gen}
\end{align}
where $  \phi  $ denotes the standard $  SU(2)_L  $ Higgs doublet, while $  \chi  $ and $  \xi  $ represent the complex and real scalar triplets, respectively, expressed in the $  SU(2)_L  $ bi-doublet representation:
\begin{align}
\phi &=
\begin{pmatrix}
\phi^+ \\
\phi^0
\end{pmatrix}, \qquad
\chi =
\begin{pmatrix}
\frac{\chi^+}{\sqrt{2}} & \chi^{++} \\
\chi^0 & -\frac{\chi^+}{\sqrt{2}}
\end{pmatrix}, \qquad
\xi =
\begin{pmatrix}
\frac{\xi^0}{\sqrt{2}} & \xi^+ \\
\xi^- & -\frac{\xi^0}{\sqrt{2}}
\end{pmatrix}.
\label{par}
\end{align}

The correspondence between the quartic couplings in the custodial-symmetric
potential (\ref{potential}) and those appearing in Eq.(\ref{pot_gen}) is
\begin{align}
\lambda &= 4\lambda_1, &
\rho_1 &= 4\lambda_2 + 6\lambda_3, &
\rho_2 &= -4\lambda_3, &
\rho_3 &= 2(\lambda_2 + \lambda_3), \nonumber \\
\rho_4 &= 4\lambda_2, &
\rho_5 &= 4\lambda_3, &
\sigma_1 &= 4\lambda_4 - \lambda_5, &
\sigma_2 &= 2\lambda_5, \nonumber \\
\sigma_3 &= 2\lambda_4, &
\sigma_4 &= \sqrt{2}\,\lambda_5.
\label{rel}
\end{align}

In the custodial-symmetric limit, one has
\begin{equation}
\lambda_{8a} = \lambda_{8b} \equiv \lambda_8.
\end{equation}

The RGEs of the general scale-invariant GM model are computed up to two-loop
order using the \texttt{PyR@TE} package \cite{sartore2021pyr}. For brevity, only the one-loop beta functions of the model couplings are presented in Appendix~D.

For the boundary conditions, we use the NNLO values of the gauge couplings and
the top Yukawa coupling evaluated at the top-quark mass scale $M_t$ \cite{buttazzo2013investigating}:
\begin{equation}
g_Y(M_t) = 0.35745, \qquad
g_2(M_t) = 0.64779, \qquad
g_3(M_t) = 1.1666, \qquad
y_t(M_t) = 0.93690.
\end{equation}

The initial values of the quartic couplings are taken at the
Gildener--Weinberg scale $\mu_{\text{GW}}$, from which the RGEs are solved.
Custodial symmetry is assumed to be conserved at the weak scale. A detailed
discussion of the impact of high-scale perturbativity is presented in the
numerical analysis.

\subsection{ Oblique Parameters}
In the scale-invariant extension of the Georgi–Machacek  model, electroweak precision observables (EWPO) provide stringent constraints on the extended scalar sector, particularly through the oblique parameters $  S  $ and $  T  $\cite{peskin1992estimation}. These parameters capture new physics contributions to gauge boson self-energies. In the standard GM model, custodial $SU(2) _L \times   SU(2)_R  $ symmetry protects the $  \rho  $ parameter at tree level, but loop-level corrections introduce deviations parameterized by $  S  $ and $  T  $.
The one-loop contributions to $  S  $ and $  T  $ arise from loops involving the additional scalar multiplets (the real and complex triplets) and the gauge-singlet scalar $  S  $, as well as mixing with the SM-like Higgs.  The $S$ parameter is defined in the usual way as
\begin{equation}
S = \frac{4 s_W^2 c_W^2}{\alpha_{\rm em} M_Z^2} \Bigl[
\Pi_{ZZ}(M_Z^2) - \Pi_{ZZ}(0)
- \frac{c_W^2 - s_W^2}{s_W c_W} \Pi_{Z\gamma}(M_Z^2)
- \Pi_{\gamma\gamma}(M_Z^2)
\Bigr],
\label{eq:Sparam}
\end{equation}
where $s_W = \sin\theta_W$, $c_W = \cos\theta_W$, $\theta_W$ is the weak mixing angle, and $\alpha_{\rm em}$ is the fine-structure constant evaluated at the $Z$ pole.

The new-physics contribution arising in the SIGM model is given by
\begin{align}
\Delta S &\equiv S_{\text{SIGM}} - S_{\text{SM}} \notag \\[1.2ex]
&= \frac{s_W^2 c_W^2}{\alpha_{\rm em}} \Biggl\{
\frac{1}{96\pi^2} \Bigl[
   |g_{ZhH_3}|^2 \, G_1(m_h^2,m_{H_3}^2)
+ |g_{ZsH_3}|^2 \, G_1(m_s^2,m_{H_3}^2) \notag \\
&\qquad\qquad\qquad
+ |g_{ZHH_3}|^2 \, G_1(m_H^2,m_{H_3}^2)
+ \bigl(|g_{ZH_3H_5}|^2 + 2|g_{ZH_3^+H_5^-}|^2\bigr) G_1(m_{H_3}^2,m_{H_5}^2)
\Bigr] \notag \\
&\qquad + \frac{1}{32\pi M_Z^2} \Biggl[
   |g_{ZZh}|^2 G_2(m_h^2,M_Z^2)
+ |g_{ZZs}|^2 G_2(m_s^2,M_Z^2)
+ |g_{ZZH}|^2 G_2(m_H^2,M_Z^2) \notag \\
&\qquad\qquad\qquad
+ |g_{ZZH_5}|^2 G_2(m_{H_5}^2,M_Z^2)
+ 2|g_{ZW^+H_5^-}|^2 G_2(m_{H_5}^2,M_W^2) \notag \\
&\qquad\qquad\qquad
- |g_{ZZh}^{\text{SM}}|^2 G_2(m_h^2,M_Z^2)
\Biggr] \notag \\
&\qquad - \frac{1}{12\pi} \Bigl( \ln m_{H_3}^2 + 5\ln m_{H_5}^2 \Bigr)
\Biggr\}.
\label{eq:DeltaS}
\end{align}

The loop functions $G_1$ and $G_2$ are defined as
\begin{equation}
G_1(m_1^2,m_2^2) = 2 + \ln m_1^2 + \ln m_2^2 + G(m_1^2,m_2^2), \qquad
G_2(m_1^2,m_2^2) = \tilde{G}(m_1^2,m_2^2),
\end{equation}
where the functions $G$ and $\tilde{G}$ are the standard finite pieces of the oblique self-energy functions, as given in Ref.~\cite{grimus2008oblique} (see also Appendix C for a summary). The couplings $g_{V_i h_j h_k}$ and $g_{V_i V_j h_k}$, with $V_i = W, Z$, parameterize the interactions of the electroweak gauge bosons with pairs of scalar mass eigenstates and with a single scalar, respectively. These couplings encode the modifications induced by scalar mixing in the extended sector and are listed explicitly in Appendix~B.

In the GM model, the one-loop contribution to the $T$ parameter is well known to be quadratically divergent, a structural feature driven by hypercharge $U(1)_Y$ gauge interactions that explicitly break the model's custodial $SU(2)_L\times SU(2)_R$ symmetry~\cite{gunion1991naturalness,englert2013triplet}. This ultraviolet sensitivity directly dictates its phenomenological treatment within our framework. Unlike the $S$ parameter, which is UV finite and can therefore be unambiguously computed from the low-energy spectrum, the quadratic divergence in $T$ must be absorbed into a custodial-symmetry-breaking counterterm. The finite part of this counterterm represents an independent boundary condition that depends entirely on the matching to the unknown UV completion entering at (or before) the cutoff scale of the effective theory.

Consequently, even when considering regions of the parameter space that remain perturbatively valid up to an intermediate high-energy scale---as will be demonstrated by the global numerical scans in Section 4, where the Landau pole typically occurs at $\mu_{\rm LP} \sim 10^9$~GeV---the $T$ parameter does not become uniquely predictable within the low-energy effective theory. Any specific choice for this finite boundary condition would be inherently scheme-dependent. For this reason, following the standard treatment adopted in the Georgi--Machacek literature, we marginalize over the $T$ parameter in the electroweak precision fit and constrain the model primarily through the ultraviolet-finite $S$ parameter. This treatment reflects the intrinsic ultraviolet sensitivity of the $T$ parameter rather than any incompleteness of the low-energy analysis.

In practice, this marginalization is implemented following the conservative procedure of \cite{hartling2015indirect}. The $\chi^2$ function for  $S$ and $T$ parameters is  
\be
\chi^2 = \frac{1}{1 - \rho_{ST}^2}
\left[
\frac{(S - S_{\mathrm{exp}})^2}{(\Delta S_{\mathrm{exp}})^2}
+
\frac{(T - T_{\mathrm{exp}})^2}{(\Delta T_{\mathrm{exp}})^2}
-
\frac{2\rho_{ST}(S - S_{\mathrm{exp}})(T - T_{\mathrm{exp}})}
{\Delta S_{\mathrm{exp}} \Delta T_{\mathrm{exp}}}
\right].
\ee
where the experimental inputs are $  S^{\rm exp} = 0.06 \pm 0.09  $, $  T^{\rm exp} = 0.10 \pm 0.07  $, $  \Delta S^{\rm exp} = 0.09  $, $  \Delta T^{\rm exp} = 0.07  $, and the correlation $  \rho_{ST} = +0.91$ \cite{gfitter2014global}. Marginalizing over $  T  $ in the $  \chi^2  $ fit to EWPO. The global $  \chi^2  $ is minimized by solving $  \partial \chi^2 / \partial T = 0  $, yielding the effective $  T  $ value:
\be
T \equiv T_{\rm min} = T^{\rm exp} + \rho_{ST} \frac{(S - S^{\rm exp}) \Delta T^{\rm exp}}{\Delta S^{\rm exp}}
\ee
 this marginalization effectively decouples $  T  $, reducing the constraint to one primarily on $  S  $.

\subsection{Higgs-boson Couplings and Signal Strengths}
\label{sec:higgscouplings}

Any extension of the SM containing additional scalar degrees of freedom must be confronted with the precise measurements of Higgs-boson properties at the LHC. We therefore compute the modified Higgs couplings induced by mixing between the SM-like Higgs field and the additional scalars of the SIGM model, and compare the resulting predictions with the latest experimental constraints.

We define the coupling modifiers as
\begin{equation}
\kappa_V = \frac{g_{hVV}^{\text{SIGM}}}{g_{hVV}^{\text{SM}}}, \qquad
\kappa_f = \frac{g_{hff}^{\text{SIGM}}}{g_{hff}^{\text{SM}}},
\label{eq:kappadef}
\end{equation}
corresponding respectively to Higgs couplings to vector bosons ($V=W,Z$) and to fermions ($f$).

In the framework of the scale-invariant Georgi-Machacek model, these quantities take the explicit form
\begin{subequations}
\begin{align}
\kappa_V &= c_H c_\alpha c_\beta
- \sqrt{\frac{8}{3}}\, s_H \bigl( c_\beta c_\gamma +c_\beta s_\alpha s_\gamma \bigr),
\label{eq:kappaV} \\
\kappa_f &= \frac{c_\alpha c_\beta}{c_H},
\label{eq:kappaf}
\end{align}
\end{subequations}
where we use the shorthand notation $c_i\equiv\cos i$ and $s_i\equiv\sin i$ for $i=\alpha,\beta,\gamma$, and the additional mixing angle $\theta_H$ is defined via
\begin{equation}
\tan\theta_H = \frac{\sqrt{8}\,v_\Delta}{v_h}, \qquad
s_H = \sin\theta_H, \qquad c_H = \cos\theta_H.
\label{eq:thetaH}
\end{equation}

In the absence of invisible or undetected decay modes, the partial decay width of the Higgs boson into any final state $XX$ is modified according to
\begin{equation}
\Gamma^{\text{SIGM}}(h\to XX) = \kappa_X^2 \, \Gamma^{\text{SM}}(h\to XX),
\label{eq:decaymod}
\end{equation}
leading to a total Higgs decay width
\begin{equation}
\Gamma_{\text{total}} = \sum_{X} \kappa_X^2 \, \Gamma^{\text{SM}}(h\to XX).
\label{eq:GammaTotal}
\end{equation}

For reference, the most important SM partial decay widths are collected below:
\begin{subequations}
\begin{align}
\Gamma(h\to f\bar{f}) &= \frac{N_c m_f^2 M_h}{8\pi v^2} \left(1-\frac{4m_f^2}{M_h^2}\right)^{3/2},
\label{eq:hff} \\
\Gamma(h\to WW) &= \frac{\pi\alpha M_h^3}{8 M_W^2 s_W^2} \cdot
\frac{1}{\sqrt{1-4x_W^2}} \Bigl(1-4x_W^2+12x_W^4\Bigr), \\
\Gamma(h\to ZZ) &= \frac{\pi\alpha M_h^3}{16 M_W^2 s_W^2 c_W^2} \cdot
\frac{1}{\sqrt{1-4x_Z^2}} \Bigl(1-4x_Z^2+12x_Z^4\Bigr),
\end{align}
\end{subequations}
with $x_W = M_W/M_h$ and $x_Z = M_Z/M_h$.

The experimentally measured signal strength in the channel $XX$ is defined as
\begin{equation}
\mu_{XX} = \frac{\sigma(pp\to h)_{\text{SIGM}}}{\sigma(pp\to h)_{\text{SM}}} \times
\frac{\text{BR}(h\to XX)_{\text{SIGM}}}{\text{BR}(h\to XX)_{\text{SM}}}.
\label{eq:muXXdef}
\end{equation}
Within the narrow-width approximation and assuming SM-like production (mainly via gluon fusion and vector-boson fusion), this can be expressed as
\begin{equation}
\mu_{XX} \simeq \kappa_f^2 \kappa_V^2 \frac{\Gamma_h^{\text{SM}}}{\Gamma_{\text{total}}},
\label{eq:muSimplified}
\end{equation}
where $\Gamma_h^{\text{SM}} \simeq 4.08 \times 10^{-3}\,\text{GeV}$ is the total SM Higgs width \cite{heinemeyer2013handbook}.

When the invisible decay $h\to ss$ is kinematically open ($m_s < m_h/2$), an additional contribution must be included:
\begin{equation}
\Gamma_{\text{total}} = \sum_{X} \kappa_X^2 \Gamma^{\text{SM}}(h\to XX) + \Gamma(h\to ss),
\label{eq:GammaTotal_invisible}
\end{equation}
with
\begin{equation}
\Gamma(h\to ss) = \frac{(g_{hss})^2}{32\pi m_h} \sqrt{1 - \frac{4m_s^2}{m_h^2}}.
\label{eq:Ghss}
\end{equation}

The trilinear coupling $g_{hss}$, obtained from the scalar potential after rotation to the mass eigenbasis, can be expressed in a compact form by 
\begin{align}
g_{hss} ={}&
12\lambda_1\, a_1 a_2^2 v_h \notag \\
&+ \lambda_2\, \mathcal{F}_2(a_i,b_i; v_h, v_\Delta)
+ 12(\lambda_3+\lambda_4)\, b_1 b_2^2 v_\Delta \notag \\
&+ \lambda_5\, \mathcal{F}_5(a_i,b_i; v_h, v_\Delta)
+ 12\lambda_6\, v_s c_\alpha c_\beta s_\alpha^2 \notag \\
&+ \lambda_7\, \mathcal{F}_7(a_i; v_h, v_s)
+ 3\lambda_8\, \mathcal{F}_8(b_i; v_\Delta, v_s),
\end{align}
where, for convenience, we have introduced the auxiliary functions
\begin{align}
\mathcal{F}_2 &= 2 v_h a_1 b_2^2 + 4 v_h b_1 a_2 b_2
+ 4 v_\Delta a_1 a_2 b_2 + 2 v_\Delta b_1 a_2^2, \notag \\
\mathcal{F}_5 &= v_h a_2 b_2^2 + 2 v_h b_1 a_2 b_2
+ 2 v_\Delta a_1 a_2 b_2 + v_h b_1 a_2^2, \notag \\
\mathcal{F}_7 &= 2 v_s c_\alpha c_\beta a_2^2 + 4 v_s a_1 s_\alpha a_2
+ 4 v_h c_\alpha c_\beta s_\alpha a_2 + 2 v_h a_1 s_\alpha^2, \notag \\
\mathcal{F}_8 &= 2 v_s c_\alpha c_\beta b_2^2 + 4 v_s b_1 s_\alpha b_2
+ 4 v_\Delta c_\alpha c_\beta s_\alpha b_2 + 2 v_\Delta b_1 s_\alpha^2.
\end{align}
where we use the shorthand notation $a_i = (\mathcal{R})_{i2}$ and $b_i = (\mathcal{R})_{i3}$, with $\mathcal{R}$ denoting the rotation matrix introduced earlier.

Having established the full set of theoretical limits (unitarity, vacuum stability, perturbativity) and experimental bounds (oblique parameters, Higgs signal strengths), we now turn to a numerical exploration of the parameter space. In the next section, we perform a systematic scan, apply the constraints derived here, and identify regions consistent with all requirements. We also examine the high-scale behavior of the couplings and the resulting implications for the model’s validity up to the Planck scale.

\section{Numerical Analysis}

The model under consideration contains nine free parameters: the eight quartic couplings, \( \lambda_1, \dots, \lambda_8 \), and the right-handed neutrino Yukawa coupling \( y_N \). This number is reduced to seven by identifying the lighter CP-even scalar \( h_1 \) with the SM-like Higgs boson observed at the LHC, \( m_h \simeq 125\,\mathrm{GeV} \), and by imposing the electroweak vacuum expectation value relation \( v^2 = v_h^2 + 8 v_\Delta^2 = (246\,\mathrm{GeV})^2 \). To further simplify the analysis, we fix the right-handed neutrino mass to \( m_N = 200\,\mathrm{GeV} \). Rather than working directly with the quartic couplings, we use the more physically transparent parameter set
\[
(m_{H_3},\, m_{H_5},\, m_H,\, \sin\beta,\, v_\Delta,\, v_s)\,.
\]
For the numerical scan, we adopt the following ranges:
$$\begin{aligned}
&m_{H_3} \in [80, 600]\,\mathrm{GeV},\quad m_{H_5} \in [84, 1000]\,\mathrm{GeV},\quad m_H \in [125, 3000]\,\mathrm{GeV},\quad \beta \in \left[-\frac{\pi}{2}, \frac{\pi}{2}\right],\\
&v_\Delta \in [0, 86.97]\,\mathrm{GeV},\quad v_s \in [0, 5000]\,\mathrm{GeV}.
\end{aligned}$$

The lower bound on \(m_{H_3}\) is taken from the latest LEP search for singly charged Higgs bosons~\cite{navas2024review,sopczak2002complete}, while the lower bound on \(m_{H_5}\) is obtained from the reinterpretation by Kanemura \textit{et al.}~\cite{Kanemura:2014ipa} of the ATLAS same-sign dilepton search at \(\sqrt{s}=8\)~TeV for doubly charged Higgs bosons decaying predominantly through \(H^{\pm\pm}\rightarrow W^{\pm(*)}W^{\pm(*)}\). The upper bounds follow from imposing perturbativity, namely \( \lambda_3, \lambda_5 < 4\pi \). To ensure that the unitarity and bounded-from-below conditions derived earlier are satisfied, we determine the quartic couplings \( \lambda_i \) for each sampled point by numerically solving a system of eight equations. This system consists of the three minimization conditions in Eq.~(\ref{mc}), the requirement \( \det(A)=0 \) for the coefficient matrix \( A \) of the linear system, and a set of relations derived from the scalar mass matrices.

\begin{align}
m_h^2 + m_H^2 &= \operatorname{Tr}(M_0^2), \notag\\
m_h^2 m_H^2 &= \frac{1}{2}\left[\bigl(\operatorname{Tr}(M_0^2)\bigr)^2 - \operatorname{Tr}\!\left((M_0^2)^2\right)\right], \notag\\
M_{H_3}^2 &= -\frac{1}{2} v_{\varphi}^2 \lambda_5\, n_h^2, \notag\\
M_{H_5}^2 &= v_{\varphi}^2\left(8\lambda_3\, n_{\Delta}^2 - \frac{3}{2}\lambda_5\, n_h^2\right).
\label{masseq}
\end{align}

To incorporate electroweak precision constraints, we compute the oblique parameter \(S\) and evaluate the corresponding chi-squared function,
\be
\chi_S^2 = \frac{(S_{\rm model} - S_{\rm exp})^2}{(\Delta S)^2}\,.
\ee
We also construct a chi-squared function for the Higgs signal strengths in five decay channels: \(h \to \mu^+\mu^-\), \(h \to b\bar{b}\), \(h \to \tau^+\tau^-\), \(h \to WW^*\), and \(h \to ZZ^*\). For each channel, the signal strength \(\mu_i\) is computed using the coupling modifiers defined in Eq.~(\ref{eq:kappaV},\ref{eq:kappaf}), and we define
\be
\chi_{\mu}^2 =
\sum_{i,j}
(\mu_i - \hat{\mu}_i)
\,(V^{-1})_{ij}\,
(\mu_j - \hat{\mu}_j)\,,
\ee
where \(\mu_i\) denotes the model prediction for channel \(i\), \(\hat{\mu}_i\) is the corresponding experimental central value, and \((V^{-1})_{ij}\) is the inverse covariance matrix. The most recent \(1\sigma\) values from the Particle Data Group~\cite{particle2022review} are
\[
\hat{\mu}_{WW} = 1.19 \pm 0.12,\quad
\hat{\mu}_{ZZ} = 1.01 \pm 0.07,\quad
\hat{\mu}_{b\bar{b}} = 0.98 \pm 0.12,\quad
\hat{\mu}_{\mu^+\mu^-} = 1.19 \pm 0.34,
\]
and 
\[
\hat{\mu}_{\tau^+\tau^-} = 1.15^{+0.16}_{-0.15}\,.
\]

\begin{figure}[H]
  \centering
  \resizebox{0.5\textwidth}{!}{\includegraphics{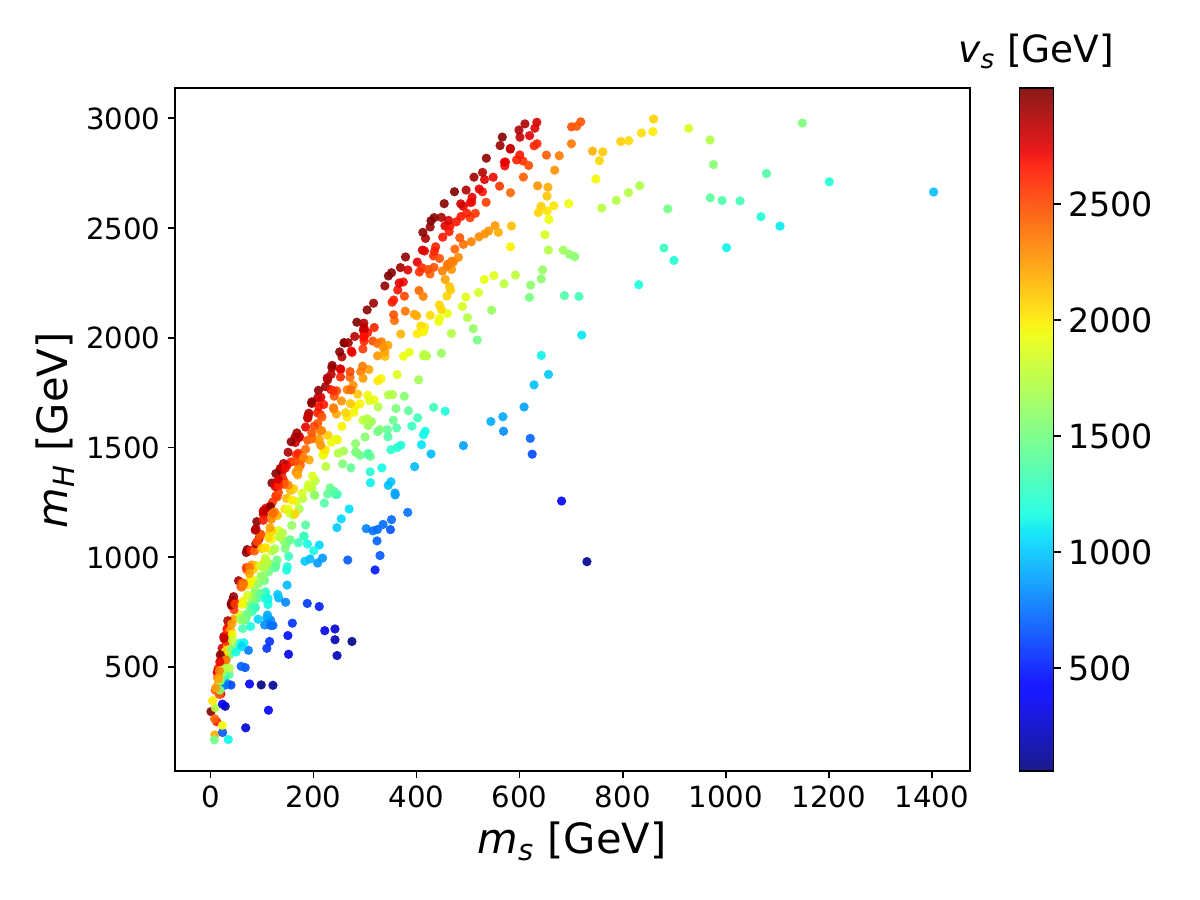}}
  \caption{Scatter plot of the allowed parameter space in the plane ($m_s,m_H$). The color coding represents the vacuum expectation value of the singlet scalar, $v_s$, in GeV. All points satisfy theoretical constraints from perturbative unitarity and vacuum stability, as well as experimental bounds from electroweak precision observables (the $S$ parameter) and LHC Higgs signal strength measurements and LHC search for doubly charged Higgs $H^{\pm\pm}$.}
  \label{msmh}
\end{figure}

These measurements place stringent constraints on the coupling modifiers $\kappa_f$ and $\kappa_V$, and therefore on the mixing angles $\alpha$, $\beta$, and $\gamma$. Figure.~\ref{msmh} shows the allowed region in the $(m_s, m_H)$ plane after imposing all theoretical constraints from perturbative unitarity and vacuum stability, together with the experimental bounds from the oblique parameters, the LHC Higgs signal strengths, and the LHC search for $H_5^{\pm\pm}$ production in vector boson fusion with decay to $W^{\pm}W^{\pm}$, with the color coding indicating different values of the singlet VEV $v_s$.

We find that the scalon mass \(m_s\) depends strongly on \(v_s\): for \(v_s \approx 1500\,\mathrm{GeV}\), it can reach values as large as \(1200\,\mathrm{GeV}\), whereas for \(v_s \approx 3000\,\mathrm{GeV}\), its upper bound decreases to about \(600\,\mathrm{GeV}\). By contrast, the heavy scalar mass \(m_H\) is less restricted and can extend up to \(3000\,\mathrm{GeV}\). Throughout the scanned parameter space, \(m_H > m_s\), which reflects the fact that the scalon mass arises radiatively, while \(m_H\) is generated already at tree level.

\begin{figure}[H]
  \centering
  \resizebox{0.45\textwidth}{!}{\includegraphics{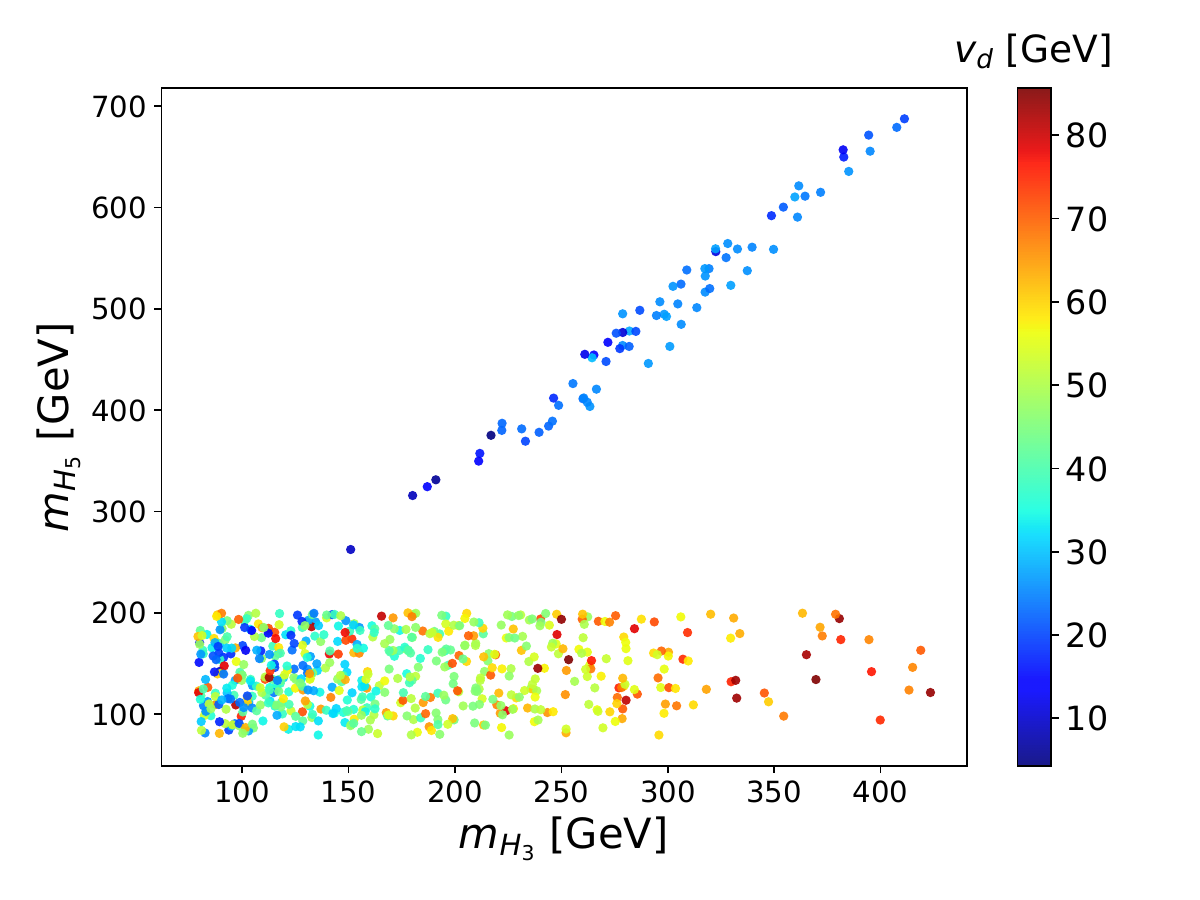}}
  \caption{Scatter plot of the allowed parameter space in the plane ($m_{H3},m_{H5}$). The color coding represents the vacuum expectation value of the triplet scalars, $v_{\Delta}$, in GeV. All points satisfy theoretical constraints from perturbative unitarity and vacuum stability, as well as experimental bounds from electroweak precision observables (the $S$ parameter) and LHC Higgs signal strength measurements and LHC search for doubly charged Higgs $H^{\pm\pm}$.}
  \label{mh3mh5}
\end{figure}

Figure~\ref{mh3mh5} shows the allowed region in the $(m_{H_3}, m_{H_5})$ plane after 
imposing the full set of theoretical and experimental constraints discussed above. The color coding indicates the profiled 
values of the triplet VEV $v_\Delta$.

Two qualitatively different regimes are visible. For $m_{H_5} \lesssim 200\,\mathrm{GeV}$, 
where the direct collider search for  $H_5^{\pm\pm}$ production via VBF with subsequent decay to $W^{\pm}W^{\pm}$ does not yet apply, the parameter space is governed by 
perturbative unitarity, vacuum stability, and the $S$ parameter  and the higgs signal measurements alone. This produces a dense 
and broadly populated region across the  range $m_{H_3} \simeq 
80$--$430\,\mathrm{GeV}$, accommodating the complete spectrum of triplet VEV values from 
a few GeV up to $v_\Delta \simeq 85\,\mathrm{GeV}$. For $m_{H_5} \gtrsim 200\, \mathrm{GeV}$
a clear near-linear correlation between $m_{H_3}$ and $m_{H_5}$ is observed, for smaller $v_\Delta$ (blue points) reflecting 
the tight constraint imposed by the $S$ parameter  and unitarity on the mass splitting within the custodial 
multiplets. As $v_\Delta$ increases, this correlation progressively weakens, allowing a 
wider range of mass splittings.

Overall, the scan reveals that the model admits two distinct phenomenological scenarios: 
a light quintet regime ($m_{H_5} \lesssim 200\,\mathrm{GeV}$) with essentially unconstrained 
$v_\Delta$ and a heavier quintet regime where a small triplet VEV is required by the combination of electroweak 
precision observables, unitarity and  LHC searches.

To assess the validity of the scale-invariant GM model at higher energy scales, we solve the two-loop renormalization group equations for the quartic couplings, as described in Section 3.3, and determine the energy scale at which each parameter point first violates the perturbativity bound \( |\lambda_i| < 4\pi \) or develops a Landau pole. The integration is terminated immediately at the first such occurrence, and the corresponding scale \( \mu_{\rm LP} \) is taken as the UV cutoff for that point.

The results are shown in Figures 3 and 4. A comprehensive scan of the viable parameter space, after imposing the theoretical and experimental constraints mentioned previously, shows that every allowed parameter point develops a Landau pole at or below approximately \(10^9\,\mathrm{GeV}\). This value therefore represents the maximal UV cutoff found in the viable parameter space, and it is attained only by points lying close to the lower boundary of the allowed scalar-mass ranges. The running of representative quartic couplings for a typical point that remains perturbative up to \(10^9\,\mathrm{GeV}\) is displayed in Figure 5, while the corresponding weak-scale input values are listed in Table 1.

\begin{table}[h!]
\centering
\caption{Benchmark parameter point surviving perturbativity up to
$\mu_{\rm LP} \sim 10^9$~GeV, after imposing all theoretical constraints
(perturbative unitarity, vacuum stability) and experimental bounds
(oblique $S$ parameter, LHC Higgs signal strengths and direct search for $H_5^{\pm\pm}$ production in VBF with decay to $W^{\pm}W^{\pm}$).
The upper block lists the physical input parameters; the middle block
gives the custodial-basis quartic couplings $\lambda_i$; the lower block
gives the general-basis couplings related to the $\lambda_i$ via
Eq.~(\ref{rel}), which enter directly the RGEs.}
\label{tab:benchmark}
\renewcommand{\arraystretch}{1.3}

\begin{tabular}{lc lc}
\toprule
\multicolumn{4}{c}{\textbf{Physical Parameters}} \\
\midrule
Parameter & Value & Parameter & Value \\
\midrule
$m_{H}$~[GeV]   & $1270.58$ & $v_s$~[GeV]        & $1763.26$ \\
$m_{H_3}$~[GeV] & $80.08$  & $v_{\Delta}$~[GeV] & $71.69$   \\
$m_{H_5}$~[GeV] & $100.95$  & $\sin\beta$        & $-0.5$    \\
$m_s$~[GeV]     & $180.22$ & $\log_{10}(\mu_{\rm LP}/\text{GeV})$ & $9.49$ \\
\midrule
\multicolumn{4}{c}{\textbf{Custodial-Basis Quartic Couplings}} \\
\midrule
$\lambda_1$ & $\phantom{-}0.08225$ & $\lambda_5$ & $-0.21194$ \\.                                
$\lambda_2$ & $\phantom{-}-0.03998$ & $\lambda_6$ & $\phantom{-} 0.02195$ \\
$\lambda_3$ & $ 0.09788$           & $\lambda_7$ & $ -0.08473$ \\
$\lambda_4$ & $\phantom{-} 0.14124$ & $\lambda_8$ & $\phantom{-}0.07323$ \\
\midrule
\multicolumn{4}{c}{\textbf{General-Basis Quartic Couplings}} \\
\midrule
$\lambda$  & $\phantom{-}0.3290$ & $\rho_4$   & $\phantom{-}-0.15992$ \\
$\rho_1$   & $\phantom{-}0.42736$ & $\rho_5$   & $0.39152$           \\
$\rho_2$   & $\phantom{-}-0.39152$ & $\sigma_1$ & $\phantom{-}0.77690$ \\
$\rho_3$   & $\phantom{-}0.11680$ & $\sigma_2$ & $-0.42388$           \\
$\sigma_3$ & $\phantom{-}0.28248$ & $\sigma_4$ & $-0.29973$           \\
\bottomrule
\end{tabular}
\end{table}

Figure~3 displays the distribution of the UV cutoff scale in the \((m_s, m_H)\) plane, with the color axis representing \(\log_{10}(\mu_{\rm LP}/{\rm GeV})\). We observe that larger values of \(m_s\) generally correspond to Landau poles appearing at lower energy scales. Figure~4 shows the analogous distribution in the \((m_{H_3}, m_{H_5})\) plane, where the cutoff scale decreases as the triplet masses increase. The highest cutoff, \(\mu_{\rm LP}\sim 10^9\,\mathrm{GeV}\), is attained only for points with \(m_{H_3} \lesssim 100\,\mathrm{GeV}\) and \(m_{H_5} \lesssim 100\,\mathrm{GeV}\).

The fact that Landau poles universally appear below \(\sim 10^9\,\mathrm{GeV}\), a scale lower than that found in the conventional GM model~\cite{blasi2017effects}, throughout the viable parameter space can be understood analytically as a direct consequence of classical scale invariance. In the GM model, the scalar potential contains dimensionful trilinear terms governed by the mass parameters \(\mu_1\), \(\mu_2\), and \(\mu_3\). These terms contribute independently to the triplet scalar masses, so that the physical masses \(m_{H_3}\) and \(m_{H_5}\) receive contributions from both the trilinear and quartic sectors. As a result, for a fixed physical spectrum, the quartic couplings \(\lambda_3\) and \(\lambda_5\) can be made arbitrarily small by an appropriate choice of the trilinear parameters. In particular, one may write schematically
\[
\lambda_5 \propto \frac{M_{H_3}^2 - \mu_{\rm eff}^2}{v^2}\,,
\]
where \(\mu_{\rm eff}\) denotes an effective combination of the trilinear mass parameters. This freedom is absent in the scale-invariant extension.

In the present model, classical scale invariance forbids all dimensionful terms at tree level, and the mass relations in Eq.~(\ref{masseq}) therefore become exact proportionalities with no compensating contributions:
\[
\lambda_5 = -\frac{2M_{H_3}^2}{v_\varphi^2 n_h^2}\,,
\qquad
\lambda_3 = \frac{M_{H_5}^2}{8 n_\Delta^2 v_\varphi^2} + \frac{3\lambda_5 n_h^2}{16 n_\Delta^2}\,.
\]
These relations imply \(|\lambda_5| \sim \mathcal{O}(1\text{--}6)\) and \(|\lambda_3| \sim \mathcal{O}(1\text{--}15)\) over the phenomenologically relevant ranges \(m_{H_3} \lesssim 430\,\mathrm{GeV}\) and \(m_{H_5} \lesssim 700\,\mathrm{GeV}\), independently of the remaining free parameters. Through the mapping in Eq.~(\ref{rel}), these values correspond to \(\rho_2 = -4\lambda_3\) and \(\sigma_2 = 2\lambda_5\), which feed directly into the large positive contributions \(+6\rho_2^2\) and \(+\sigma_2^2\) in the beta functions \(\beta_{\rho_2}\) and \(\beta_\lambda\), respectively.

We therefore conclude that the relatively low UV cutoff of order \(10^9\,\mathrm{GeV}\) is not a numerical artifact of the scan, but rather a structural consequence of classical scale invariance itself. The model should therefore be regarded as an effective theory valid up to a relatively low-to-intermediate energy scale.

\begin{figure}[H]
  \centering
  \resizebox{0.4\textwidth}{!}{\includegraphics{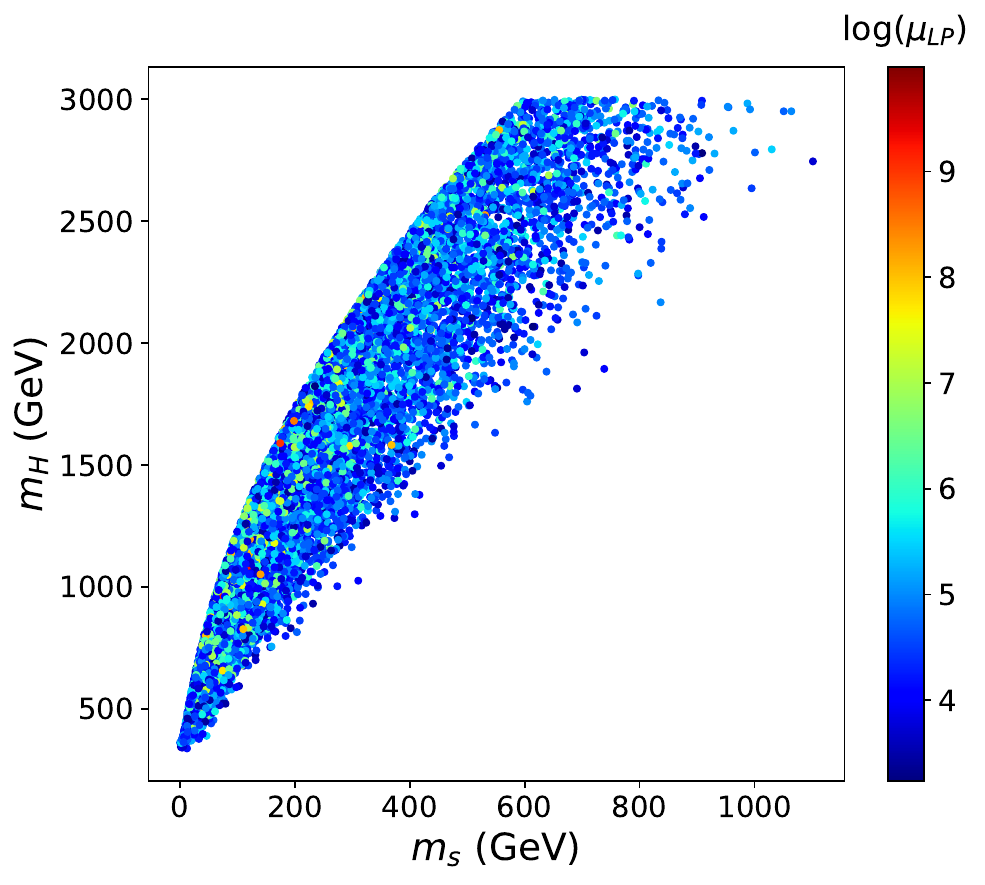}}
  \caption{Scatter plot of the allowed parameter space in the plane ($m_s,m_H$). the color code represents the log energy scale at which a landau pole or perturbativity violation appears}
  \label{msmh_cut}
\end{figure}

\begin{figure}[H]
  \centering
  \resizebox{0.4\textwidth}{!}{\includegraphics{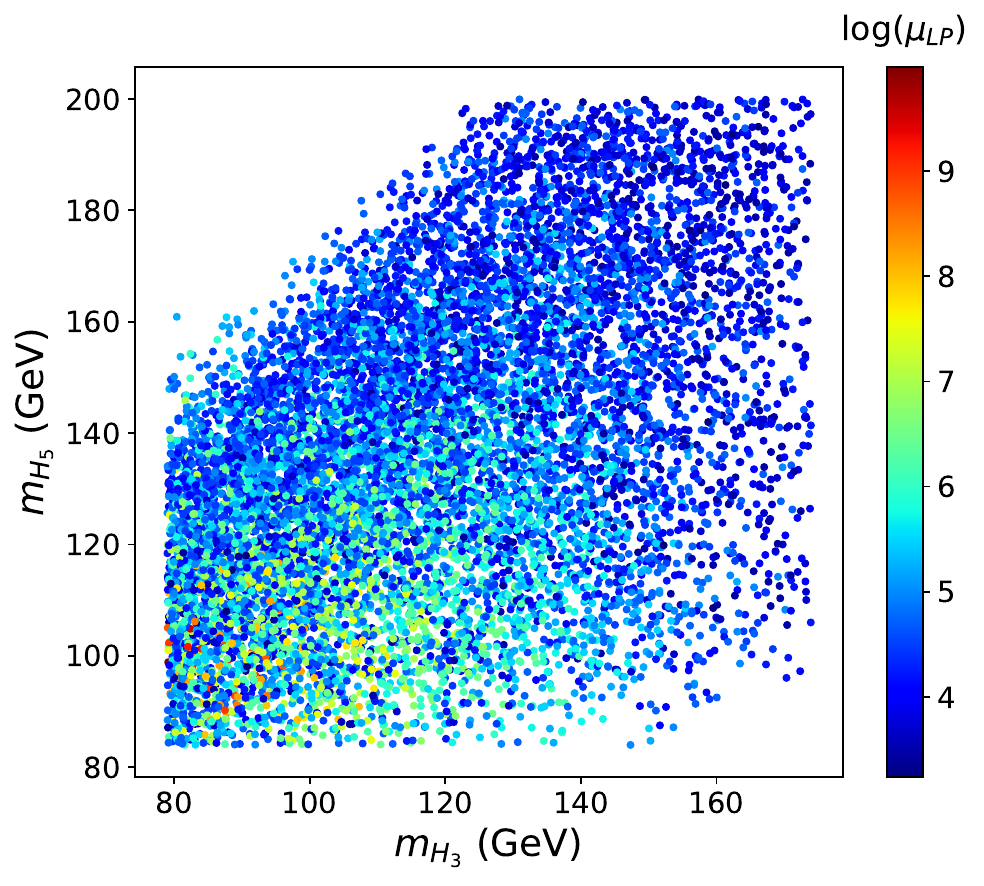}}
  \caption{Scatter plot of the allowed parameter space in the plane ($m_{H3},m_{H5}$). the color code represents the log energy scale at which a landau pole or perturbativity violation appears}
  \label{mh3mh5_cut}
\end{figure}

\begin{figure}[H]
  \centering
  \resizebox{0.5\textwidth}{!}{\includegraphics{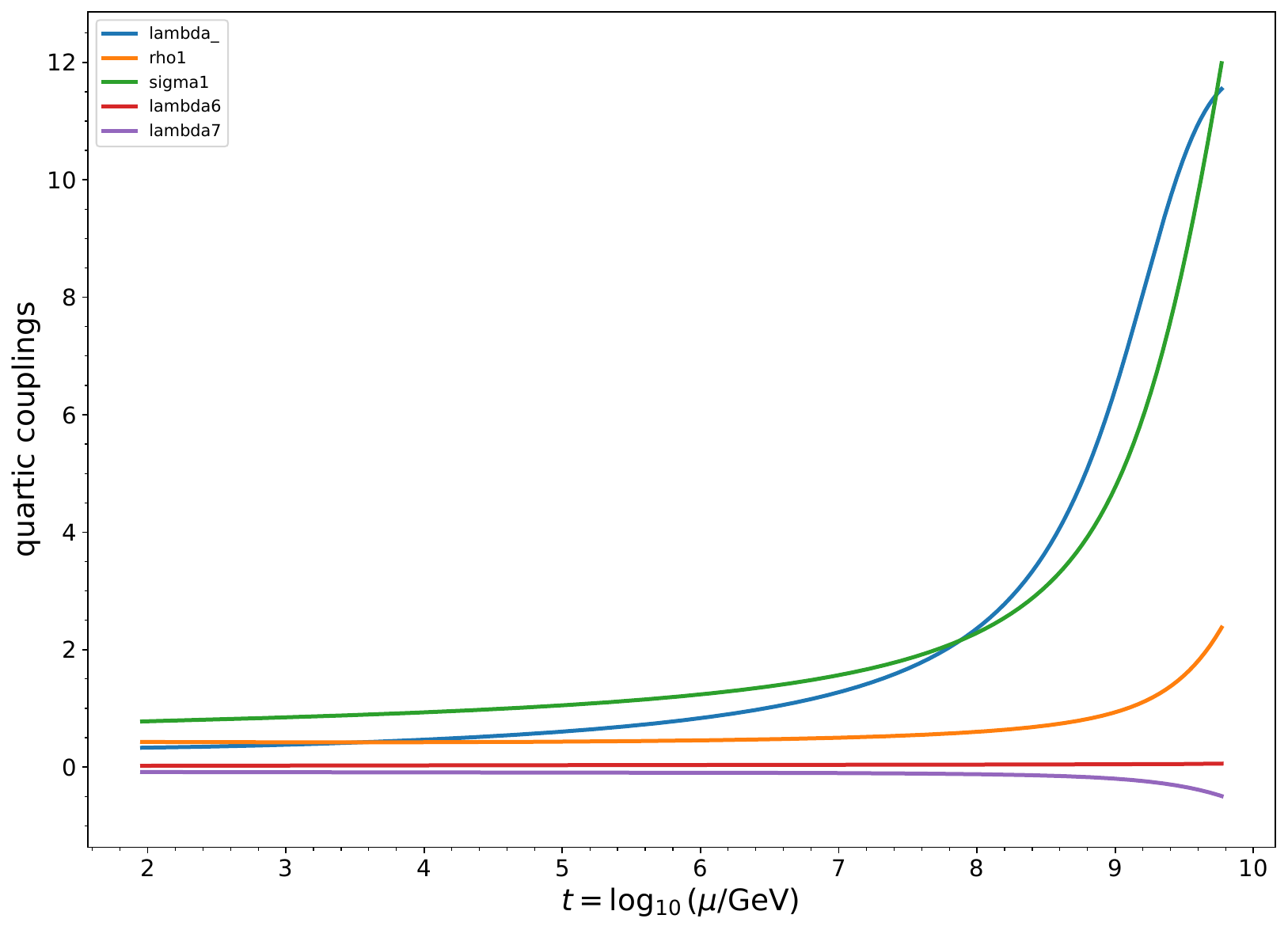}}
  \caption{Evolution of some  quartic couplings for the initial values mentioned in table1 . For this benchmark $\sigma_1$  diverges first.}
  \label{evol}
\end{figure}

\section{Summary and Conclusions}

In this work, we have constructed a classically scale-invariant extension of the Georgi--Machacek model that addresses the hierarchy problem by generating the electroweak scale dynamically through dimensional transmutation. By adding a real gauge-singlet  to the Higgs doublet and scalar triplets, and by employing the Gildener--Weinberg framework, we have shown that all mass scales can arise radiatively.

The CP-even neutral sector of the model contains three physical states: the SM-like 125 GeV Higgs boson, an additional heavy neutral scalar $H$, and a pseudo-Goldstone boson of broken scale symmetry, whose mass is generated at one loop. Custodial symmetry is preserved at tree level, protecting the $\rho$ parameter. However, the absence of the trilinear term required by scale invariance renders the type-II seesaw mechanism inoperative. To account for neutrino masses we therefore introduce a type-I seesaw via three right-handed Majorana neutrinos, with their Majorana masses generated from the singlet VEV.

We performed a comprehensive analysis of both theoretical and experimental constraints. The perturbative-unitarity conditions, as established in the literature, were implemented, while the vacuum-stability requirements were derived within our framework. In addition, high-scale perturbativity was examined through renormalization-group evolution. Electroweak precision observables, in particular the $S$ parameter, together with LHC Higgs signal-strength measurements and constraints from direct search for $H_5^{\pm\pm}$ production in VBF with decay to $W^{\pm}W^{\pm}$, impose stringent constraints on the viable parameter space.

A large numerical scan identifies significant viable regions. The scalon mass $m_s$ can reach up to $\mathcal{O}(10^3)\,\mathrm{GeV}$ for $v_s \simeq 1500\,\mathrm{GeV}$ (and has a lower upper bound for larger $v_s$), while the heavy scalar $H$ extends up to $\sim 3000\,\mathrm{GeV}$, with $m_H > m_s$ across the allowed parameter space. The custodial triplet states $H_3$ and $H_5$ populate the ranges $m_{H_3} \simeq 80$--$430\,\mathrm{GeV}$ and $m_{H_5} \simeq 84$--$700\,\mathrm{GeV}$ in two qualitatively distinct regimes: for $m_{H_5}\lesssim200\,\mathrm{GeV}$, where the direct $H_5^{\pm\pm}\to W^\pm W^\pm$ search does not yet apply, the full range of triplet VEVs up to $v_\Delta\simeq85\,\mathrm{GeV}$ remains viable. For $m_{H_5}\gtrsim200\,\mathrm{GeV}$ these same constraints combine with the direct LHC search to increasingly restrict the model to small $v_\Delta$ where a near-linear correlation between $m_{H_3}$ and $m_{H_5}$ is observed driven by the $S$ parameter and unitarity. Requiring high-scale perturbativity significantly tightens these ranges, concentrating viable points near $m_{H_3}, m_{H_5} \lesssim 100\,\mathrm{GeV}$ in the most restrictive cases.

The model's high-scale validity is strictly limited. A careful numerical analysis, tracking the first violation of $|\lambda_i| < 4\pi$ or the occurence of a Landau pole at each RGE integration step, reveals that no parameter point in the viable region remains perturbative beyond $\mu \sim 10^9$ GeV. The apparence of Landau poles indicate that the scale-invariant GM model should  be understood as a low-to-intermediate energy effective theory, valid up to $\mathcal{O}(10^9)$ GeV, requiring a UV completion at or below that scale.

Future work should include a detailed investigation of collider signatures, particularly the production and decay channels of the scalon, which may provide a distinctive probe of the model. It would also be worthwhile to explore the 
embedding of viable dark-matter candidates within this framework. In particular, the presence of a $\mathbb{Z}_2$ symmetry under which the triplet $\Delta$ is odd implies that, in the special case $v_\Delta = 0$, the neutral component of the 
triplet is stabilized and can serve as a dark-matter candidate, as previously studied in the context of the conventional GM model~\cite{lu2025dark}. It would be interesting to further examine a fully scale-invariant realization of this model.

In summary, the scale-invariant Georgi--Machacek model provides a consistent framework that addresses the hierarchy problem, accommodates small neutrino masses, and preserves the characteristic collider signatures of the GM setup. It remains compatible with current theoretical and experimental constraints and offers a promising direction for physics beyond the Standard Model.

\section*{Data Availability Statement}
No data was used for the research described in the article.
\section*{Declaration of interests}
The authors declare that they have no known competing financial interests or personal relationships that could have appeared to influence the work reported in this paper. 

\section*{Acknowledgements}
This work is supported by the Algerian Ministry of High Education and Scientific Research.

\appendix

\section{Mass matrices for charged and CP odd neutral scalars}
\subsection{Charged scalars}
Within the framework of the scale-invariant GM model, we derive the mass matrix for the charged Higgs sector using the relation
\[
M_{ij} = \frac{\partial^2 V}{\partial \varphi_i \partial \varphi_j},
\]
where \(\varphi_i\) denotes the charged scalar fields (i.e., \(\varphi_i = \zeta^{\pm}\) or \(\varphi_i = X^{\pm}\)). We find that

\begin{equation} 
\mathcal{M}_+^2 = v_{\varphi}^2 
\begin{pmatrix}
(12\lambda_4+8\lambda_3)\,n_{\Delta}^2+2\lambda_2\,n_h^2    & -4\lambda_3\,n_\delta^2+\frac{\lambda_5}{2}n_h^2 \\
-4\lambda_3\,n_\delta^2+\frac{\lambda_5}{2}n_h^2 & (12\lambda_4+8\lambda_3)\,n_{\Delta}^2+2\lambda_2\,n_h^2 
\end{pmatrix}\,,
\label{eq:massmatrix_charged}
\end{equation}
We diagonalize this matrix using the orthogonal matrix \(O\) given by
\begin{equation} 
\mathcal{O} = 
\begin{pmatrix}
\frac{1}{\sqrt{2}}    & -\frac{1}{\sqrt{2}}   \\
\frac{1}{\sqrt{2}}   & \frac{1}{\sqrt{2}}  
\end{pmatrix}\,,
\label{eq:O_charged}
\end{equation}
and we find the following mass eigenstates 
\begin{align}
M_{H_3^+}^2 &= - v_{\varphi}^2\frac{\lambda_5}{2}n_h^2,  \notag\\[1mm]
M_{H_5^+}^2 &= v_{\varphi}^2\left(8\lambda_3\,n_{\Delta}^2-\frac{3}{2}\lambda_5 n_h^2\right).
\label{eq:mass_eigenstates_charged}
\end{align}
From these results we see that the constraint \(\lambda_5<0\) must be imposed to ensure positivity of the masses. 

For the doubly charged Higgs boson, we calculate the mass as
\begin{equation}
M_{H_5^{++}}^2=\frac{\partial^2 V}{\partial X^{++} \partial X^{--}}=v_{\varphi}^2\left(8\lambda_3\,n_{\Delta}^2-\frac{3}{2}\lambda_5 n_h^2\right),
\label{eq:mass_doubly_charged}
\end{equation}
We see that $M_{H_5^{++}}^2=M_{H_5^{+}}^2$ which confirm the mass degeneracy in the quintet \(H_5\). 

\subsection{CP odd neutral Higgs}
As in the charged case, we derive the mass matrix for CP odd Higgs using
\[
M_{ij}=\frac{\partial^2 V}{\partial \varphi_i \partial \varphi_j},
\]
where in this case \(\varphi_i=\phi_i\) or \(\varphi_i=X_i\), the imaginary parts of \(X^0\) and \(\phi_0\), respectively. The mass matrix is 
\begin{equation} 
\mathcal{M}_{\text{CP odd}}^2 = v_{\varphi}^2 
\begin{pmatrix}
4\lambda_1\,n_h^2+(6\lambda_2-2\lambda_5)\,n_\delta^2+2\lambda_7\,n_s^2    & \sqrt{2}\lambda_5 n_h n_{\Delta} \\
\sqrt{2}\lambda_5 n_h n_{\Delta} & (12\lambda_2+4\lambda_3)\,n_{\Delta}^2+(2\lambda_4+\frac{\lambda_5}{2})\,n_h^2+2\lambda_8\,n_s^2 
\end{pmatrix}\,,
\label{eq:massmatrix_CPodd}
\end{equation}
Diagonalizing by using the rotation matrix 
\begin{equation} 
\mathcal{O} = 
\begin{pmatrix}
\cos\theta_H    & -\sin\theta_H \\
\sin\theta_H & \cos\theta_H 
\end{pmatrix}\,,
\label{eq:O_CPodd}
\end{equation}
 we find the physical masses
\begin{align}
M_{G^0}^2 &= 0,  \notag\\[1mm]
M_{H_3^0}^2 &= - v_{\varphi}^2\frac{\lambda_5}{2}n_h^2.
\label{eq:mass_eigenstates_CPodd}
\end{align}
Clearly, \(G^0\) is the neutral Goldstone boson associated with electroweak symmetry breaking and \(H_3^0\) is the CP odd neutral component of the \(H_3\) triplet.

\section{Gauge Higgs Couplings}
The couplings of scalars to gauge bosons originate from the gauge-kinetic terms in the Lagrangian:
\begin{equation}
\mathcal{L} \supset (D_\mu \Phi)^\dagger (D^\mu \Phi) 
+ \frac{1}{2}\,(D_\mu \xi)^\dagger (D^\mu \xi)
+ (D_\mu \chi)^\dagger (D^\mu \chi),
\tag{A4}
\end{equation}
where the scalar multiplets are defined as
\[
\Phi = \begin{pmatrix} \phi^+ \\ \phi^0 \end{pmatrix}, \quad
\xi = \begin{pmatrix} \xi^+ \\ \xi^0 \\ \xi^- \end{pmatrix}, \quad
\chi = \begin{pmatrix} \chi^{++} \\ \chi^+ \\ \chi^0 \end{pmatrix}.
\]
The covariant derivative is given by
\begin{equation}
D_\mu = \partial_\mu 
+ i\frac{g}{\sqrt{2}} \Bigl( W_\mu^+ T^+ + W_\mu^- T^- \Bigr)
+ i\frac{e}{s_W c_W}\, Z_\mu \Bigl( T_3 - s_W^2 Q \Bigr)
+ i e A_\mu Q,
\tag{A5}
\end{equation}
where \(T^+\) and \(T^-\) denote the raising and lowering operators of the \(SU(2)\) group, \(T^3\) is the third generator, and \(Q\) represents the charge operator.

\subsection{Couplings of Two Scalars to One Gauge Boson}
In our model, the couplings between two scalar fields and a single \( Z \) boson are derived as follows:

\begin{align}
g_{hH_3Z} &= -\sqrt{2}\,g_Z\left(-\frac{v_{\Delta}}{v}(-c_\beta c_\gamma s_\alpha + s_\beta s_\gamma) 
+ \frac{1}{\sqrt{3}}\frac{v_h}{v}(-c_\gamma s_\beta - c_\beta s_\alpha s_\gamma)\right), \notag \\
g_{sH_3Z} &=  -\sqrt{2}\,g_Z\left(-\frac{v_{\Delta}}{v}c_\alpha c_\gamma 
+ \frac{1}{\sqrt{3}}\frac{v_h}{v}c_\alpha s_\gamma\right), \notag \\
g_{HH_3Z} &=  -\sqrt{2}\,g_Z\left(-\frac{v_{\Delta}}{v}(-c_\gamma s_\alpha s_\beta - c_\beta s_\gamma) 
+ \frac{1}{\sqrt{3}}\frac{v_h}{v}(c_\beta c_\gamma - s_\alpha s_\beta s_\gamma)\right).
\tag{A6}
\end{align}
where, \( g_Z \equiv \frac{e}{s_W c_W} \).

\subsection{ Couplings of One Scalar to Two Gauge Bosons}
The couplings between one scalar field and two \( Z \) bosons are derived as follows:

\begin{align}
g_{hZZ} &= -g_Z^2\left(\frac{1}{4}v_h(-c_\beta c_\gamma s_\alpha + s_\beta s_\gamma) 
+ \frac{2}{\sqrt{3}}v_\Delta(-c_\gamma s_\beta - c_\beta s_\alpha s_\gamma)\right), \notag \\
g_{sZZ} &= -g_Z^2\left(\frac{1}{4}v_h c_\alpha c_\gamma 
+ \frac{2}{\sqrt{3}}v_\Delta c_\alpha s_\gamma\right), \notag \\
g_{HZZ} &= -g_Z^2\left(\frac{1}{4}v_h(-c_\gamma s_\alpha s_\beta - c_\beta s_\gamma) 
+ \frac{2}{\sqrt{3}}v_\Delta(c_\beta c_\gamma - s_\alpha s_\beta s_\gamma)\right).
\tag{A7}
\end{align}

In addition to these modifed couplings arising from the scale-invariant GM framework, we use previously calculated gauge-scalar couplings from the GM model to compute the \( S \)-parameter:
\begin{align}
g_{Z H_3^0 H_5^0} &= i \frac{1}{3} \frac{e}{s_W c_W} \frac{v_h}{v}, \notag \\
g_{Z H_3^+ H_5^+} &= \frac{e}{2 s_W c_W} \frac{v_h}{v}, \notag \\
g_{H_5^0 ZZ} &= \frac{8}{3} \frac{e^2}{s_W^2 c_W^2} v_\Delta, \notag \\
g_{H_5^+ W^+ Z} &= \frac{\sqrt{2} e^2 v}{c_W s_W^2}, \notag \\
g_{ ZZ h}^{\text{SM}} &= \frac{e^2 v}{2 s_W^2 c_W^2}.
\tag{A8}
\end{align}

\section{Loop Functions}
In this appendix, we present the loop functions used in the computation of oblique parameters. The function \(G\) is defined as
\begin{align}
G \left( m_1^2, m_2^2, q^2 \right)=&
- \frac{16}{3}
+ \frac{5 \left( m_1^2 + m_2^2 \right)}{q^2}
- \frac{2 \left( m_1^2 - m_2^2 \right)^2}{q^4}
+ \frac{3}{q^2}
\left[ \frac{m_1^4 + m_2^4}{m_1^2 - m_2^2}
- \frac{m_1^4 - m_2^4}{q^2}
+ \frac{\left( m_1^2 - m_2^2 \right)^3}{3 q^4} \right]
\ln\frac{m_1^2}{m_2^2}\notag\\
&+ \frac{r}{q^6}\, f \left( t, r \right).
\end{align}

The function \(\tilde{G}\) is given by
\[
\tilde{G} \left( m_1^2, m_2^2, q^2 \right) \equiv
- 2 + \left( \frac{m_1^2- m_2^2}{q^2} - \frac{m_1^2 + m_2^2}{m_1^2 - m_2^2} \right) \ln\frac{m_1^2}{m_2^2}
+ \frac{f \left( t, r \right)}{q^2}.
\]

Here, the parameters \(t\) and \(r\) are defined as
\[
t \equiv m_1^2 + m_2^2 - q^2, \qquad
r \equiv q^4 - 2 q^2 \left( m_1^2 + m_2^2 \right) + \left( m_1^2 - m_2^2 \right)^2.
\]

The auxiliary function \(f(t,r)\) is then defined by
\[
f \left( t, r \right) \equiv \begin{cases}
\sqrt{r}\, \ln \left|\frac{t - \sqrt{r}}{t + \sqrt{r}} \right|, & \text{if } r > 0,\\[3mm]
0, & \text{if } r = 0,\\[2mm]
2\, \sqrt{-r}\, \arctan\frac{\sqrt{-r}}{t}, & \text{if } r < 0.
\end{cases}
\]

Note that the absolute value in the argument of the logarithm is required only when
\[
q^2 > \left( m_1 + m_2 \right)^2.
\]
\
\section{One-loop renormalization-group equations}
\label{app:RGE}
In this appendix we collect the one-loop renormalization-group equations for the gauge, Yukawa and scalar couplings used in our numerical
analysis.  We define
\begin{equation}
\beta_X \equiv \frac{dX}{d\ln\mu},
\end{equation}
and all one-loop $\beta$-functions are given by
\begin{equation}
\beta_X = \frac{1}{16\pi^2}\,\beta_X^{(1)} .
\end{equation}

For compactness we introduce the trace combinations
\begin{align}
T_u   &\equiv \mathrm{Tr}(Y_u Y_u^\dagger), &
T_d   &\equiv \mathrm{Tr}(Y_d Y_d^\dagger), \nonumber\\
T_e   &\equiv \mathrm{Tr}(Y_e Y_e^\dagger), &
T_\nu &\equiv \mathrm{Tr}(Y_\nu Y_\nu^\dagger), &
T_N   &\equiv \mathrm{Tr}(Y_N Y_N^\dagger),
\end{align}
which in the numerical implementation are approximated by the dominant
eigenvalues, e.g.\ $T_u\simeq Y_t^2$, $T_N\simeq Y_N^2$.

\subsection*{Gauge couplings}

The one loop beta function for gauge couplings are:

\begin{align}
\beta_{g_1}^{(1)} &= \frac{47}{6}\,g_1^3,\\
\beta_{g_2}^{(1)} &= -\frac{13}{6}\,g_2^3,\\
\beta_{g_3}^{(1)} &= -7\,g_3^3.
\end{align}

\subsection*{Yukawa couplings}

The one loop beta function for Yukawa couplings are:

\begin{align}
\beta_{Y_t}^{(1)} &=
-\frac{17}{12}g_1^2 Y_t - \frac{9}{4}g_2^2 Y_t - 8 g_3^2 Y_t
+ 3T_d Y_t + T_e Y_t + T_\nu Y_t + 3T_u Y_t
- \frac{3}{2}Y_b^2 Y_t + \frac{3}{2}Y_t^3,
\\[4pt]
\beta_{Y_b}^{(1)} &=
-\frac{5}{12}g_1^2 Y_b - \frac{9}{4}g_2^2 Y_b - 8 g_3^2 Y_b
+ 3T_d Y_b + T_e Y_b + T_\nu Y_b + 3T_u Y_b
+ \frac{3}{2}Y_b^3 - \frac{3}{2}Y_t^2 Y_b,
\\[4pt]
\beta_{Y_\tau}^{(1)} &=
-\frac{15}{4}g_1^2 Y_\tau - \frac{9}{4}g_2^2 Y_\tau
+ 3T_d Y_\tau + T_e Y_\tau + T_\nu Y_\tau + 3T_u Y_\tau
+ \frac{3}{2}Y_\tau^3 - \frac{3}{2}Y_\nu^2 Y_\tau,
\\[4pt]
\beta_{Y_\nu}^{(1)} &=
-\frac{3}{4}g_1^2 Y_\nu - \frac{9}{4}g_2^2 Y_\nu
+ 3T_d Y_\nu + T_e Y_\nu + T_\nu Y_\nu + 3T_u Y_\nu
- \frac{3}{2}Y_\tau^2 Y_\nu
+ 2 Y_\nu Y_N^2 + \frac{3}{2}Y_\nu^3,
\\[4pt]
\beta_{Y_N}^{(1)} &=
4 T_N Y_N + 12 Y_N^3 + 2 Y_N Y_\nu^2 .
\end{align}

\subsection*{Scalar quartic couplings}

The one loop beta function for scalar quartic couplings are:

\begin{align}
\beta_{\lambda}^{(1)} &=
\frac{3}{8}g_1^4 + \frac{3}{4}g_1^2 g_2^2 - 3 g_1^2 \lambda
+ \frac{9}{8}g_2^4 - 9 g_2^2 \lambda + 24 \lambda^2
+ 12\lambda T_d + 4\lambda T_e + 4\lambda T_\nu + 12\lambda T_u \nonumber\\
&\quad
+ 8\lambda_7^2
+ 3\sigma_1^2 + 3\sigma_1\sigma_2 + \frac{5}{4}\sigma_2^2
+ 6\sigma_3^2 + 2\sigma_4^2
- 6T_d^2 - 2T_e^2 - 2T_\nu^2 - 6T_u^2 ,
\\[6pt]
\beta_{\rho_1}^{(1)} &=
6g_1^4 - 12g_1^2 g_2^2 - 12g_1^2\rho_1 + 15g_2^4 - 24g_2^2\rho_1
+ 8\lambda_{8a}^2
+ 28\rho_1^2 + 24\rho_1\rho_2 + 6\rho_2^2 \nonumber\\
&\quad
+ 6\rho_4^2 + 4\rho_4\rho_5 + 3\rho_5^2
+ 2\sigma_1^2 + 2\sigma_1\sigma_2 ,
\\[4pt]
\beta_{\rho_2}^{(1)} &=
24g_1^2 g_2^2 - 12g_1^2\rho_2 - 6g_2^4 - 24g_2^2\rho_2
+ 24\rho_1\rho_2 + 18\rho_2^2 - 2\rho_5^2 + \sigma_2^2 ,
\\[4pt]
\beta_{\rho_3}^{(1)} &=
6g_2^4 - 24g_2^2\rho_3 + 4\lambda_{8b}^2
+ 44\rho_3^2 + 6\rho_4^2 + 4\rho_4\rho_5 + 2\rho_5^2 + 4\sigma_3^2 ,
\\[4pt]
\beta_{\rho_4}^{(1)} &=
- 6g_1^2\rho_4 + 6g_2^4 - 24g_2^2\rho_4
+ 8\lambda_{8a}\lambda_{8b}
+ 16\rho_1\rho_4 + 4\rho_1\rho_5
+ 12\rho_2\rho_4 + 4\rho_2\rho_5 \nonumber\\
&\quad
+ 20\rho_3\rho_4 + 4\rho_3\rho_5
+ 8\rho_4^2 + 2\rho_5^2
+ 4\sigma_1\sigma_3 + 2\sigma_2\sigma_3 + 2\sigma_4^2 ,
\\[4pt]
\beta_{\rho_5}^{(1)} &=
- 6g_1^2\rho_5 + 6g_2^4 - 24g_2^2\rho_5
+ 4\rho_1\rho_5 + 8\rho_3\rho_5 + 16\rho_4\rho_5
+ 10\rho_5^2 - 2\sigma_4^2 .
\end{align}

\begin{align}
\beta_{\sigma_1}^{(1)} &=
3g_1^4 - 6g_1^2 g_2^2 - \frac{15}{2}g_1^2\sigma_1
+ 6g_2^4 - \frac{33}{2}g_2^2\sigma_1
+ 12\lambda\sigma_1 + 4\lambda\sigma_2
+ 16\lambda_7\lambda_{8a} \nonumber\\
&\quad
+ 16\rho_1\sigma_1 + 6\rho_1\sigma_2 + 12\rho_2\sigma_1 + 2\rho_2\sigma_2
+ 12\rho_4\sigma_3 + 4\rho_5\sigma_3
+ 4\sigma_1^2 + \sigma_2^2 + 2\sigma_4^2 \nonumber\\
&\quad
+ 6\sigma_1 T_d + 2\sigma_1 T_e + 2\sigma_1 T_\nu + 6\sigma_1 T_u ,
\\[4pt]
\beta_{\sigma_2}^{(1)} &=
12g_1^2 g_2^2 - \frac{15}{2}g_1^2\sigma_2
- \frac{33}{2}g_2^2\sigma_2
+ 4\lambda\sigma_2 + 4\rho_1\sigma_2 + 8\rho_2\sigma_2 \nonumber\\
&\quad
+ 8\sigma_1\sigma_2 + 4\sigma_2^2 + 4\sigma_4^2
+ 6\sigma_2 T_d + 2\sigma_2 T_e + 2\sigma_2 T_\nu + 6\sigma_2 T_u ,
\\[4pt]
\beta_{\sigma_3}^{(1)} &=
- \frac{3}{2}g_1^2\sigma_3 + 3g_2^4
- \frac{33}{2}g_2^2\sigma_3
+ 12\lambda\sigma_3 + 8\lambda_7\lambda_{8b}
+ 20\rho_3\sigma_3 \nonumber\\
&\quad
+ 6\rho_4\sigma_1 + 3\rho_4\sigma_2
+ 2\rho_5\sigma_1 + \rho_5\sigma_2
+ 8\sigma_3^2 + 4\sigma_4^2 \nonumber\\
&\quad
+ 6\sigma_3 T_d + 2\sigma_3 T_e + 2\sigma_3 T_\nu + 6\sigma_3 T_u ,
\\[4pt]
\beta_{\sigma_4}^{(1)} &=
- \frac{9}{2}g_1^2\sigma_4 - \frac{33}{2}g_2^2\sigma_4
+ 4\lambda\sigma_4 + 4\rho_4\sigma_4 - 2\rho_5\sigma_4
+ 4\sigma_1\sigma_4 + 4\sigma_2\sigma_4 + 8\sigma_3\sigma_4 \nonumber\\
&\quad
+ 6\sigma_4 T_d + 2\sigma_4 T_e + 2\sigma_4 T_\nu + 6\sigma_4 T_u .
\end{align}

\begin{align}
\beta_{\lambda_6}^{(1)} &=
72\lambda_6^2 + 16\lambda_6 T_N
+ 8\lambda_7^2 + 12\lambda_{8a}^2 + 6\lambda_{8b}^2
- 16T_N^2 ,
\\[4pt]
\beta_{\lambda_7}^{(1)} &=
-\frac{3}{2}g_1^2\lambda_7 - \frac{9}{2}g_2^2\lambda_7
+ 12\lambda\lambda_7 + 24\lambda_6\lambda_7
+ 16\lambda_7^2 + 8\lambda_7 T_N \nonumber\\
&\quad
+ 6\lambda_7 T_d + 2\lambda_7 T_e + 2\lambda_7 T_\nu + 6\lambda_7 T_u
+ 6\lambda_{8a}\sigma_1 + 3\lambda_{8a}\sigma_2 + 6\lambda_{8b}\sigma_3
- 8T_\nu T_N ,
\\[4pt]
\beta_{\lambda_{8a}}^{(1)} &=
- 6g_1^2\lambda_{8a} - 12g_2^2\lambda_{8a}
+ 24\lambda_6\lambda_{8a}
+ 4\lambda_7\sigma_1 + 2\lambda_7\sigma_2 \nonumber\\
&\quad
+ 16\lambda_{8a}^2 + 16\lambda_{8a}\rho_1 + 12\lambda_{8a}\rho_2
+ 8\lambda_{8a} T_N
+ 6\lambda_{8b}\rho_4 + 2\lambda_{8b}\rho_5 ,
\\[4pt]
\beta_{\lambda_{8b}}^{(1)} &=
- 12g_2^2\lambda_{8b}
+ 24\lambda_6\lambda_{8b}
+ 8\lambda_7\sigma_3
+ 12\lambda_{8a}\rho_4 + 4\lambda_{8a}\rho_5
+ 16\lambda_{8b}^2 + 20\lambda_{8b}\rho_3
+ 8\lambda_{8b} T_N .
\end{align}

\bibliographystyle{elsarticle-num}
\bibliography{phys}

\end{document}